\documentclass[twocolumn]{aastex631}
\shorttitle{Active Black Holes in DESI}
\shortauthors{Pucha et al.}
\usepackage{amssymb}
\usepackage{amsmath}

\newcommand{\ha}{\textrm{H}\ensuremath{\alpha}}
\newcommand{\hb}{\textrm{H}\ensuremath{\beta}}

\newcommand{\oi}{[\textrm{O}\,\textsc{i}]}

\newcommand{\oiii}{[\textrm{O}\textsc{iii}]}
\newcommand{\nii}{[\textrm{N}\textsc{ii}]}
\newcommand{\sii}{[\textrm{S}\textsc{ii}]}

\newcommand{\oiiilam}{[\textrm{O}\textsc{iii}]\ensuremath{\lambda}5007}
\newcommand{\niilam}{[\textrm{N}\textsc{ii}]\ensuremath{\lambda}6584}

\newcommand{\siilam}{[\textrm{S}\textsc{ii}]\ensuremath{\lambda\lambda}6717,6731}
\newcommand{\niilamlam}{[\textrm{N}\textsc{ii}]\ensuremath{\lambda\lambda}6548,6584}

\def\deg{^{\circ}}

\newcommand{\mbh}{{M}\ensuremath{_{\rm BH}}}
\newcommand{\mstar}{{M}\ensuremath{_{\star}}}
\newcommand{\msun}{{\rm M}\ensuremath{_{\odot}}}
\newcommand{\logmass}{\ensuremath{\log (\mstar/\msun)}}
\newcommand{\logmbh}{\ensuremath{\log (\mbh/\msun)}}

\newcommand{\ergscmsq}{\textrm{ergs s$^{-1}$ cm$^{-2}$}}

\begin{document}

\title{Tripling the Census of Dwarf AGN Candidates Using DESI Early Data}

\correspondingauthor{Ragadeepika Pucha}
\email{raga.pucha@utah.edu}

\suppressAffiliations

\author[0000-0002-4940-3009]{Ragadeepika~Pucha}
\affiliation{Department of Physics and Astronomy, University of Utah, 115 South 1400 East, Salt Lake City, UT 84112, USA}
\affiliation{Steward Observatory, University of Arizona, 933 North Cherry Avenue, Tucson, AZ 85721, USA}

\author[0000-0002-0000-2394]{S.~Juneau}
\affiliation{NSF NOIRLab, 950 N. Cherry Ave., Tucson, AZ 85719, USA}

\author[0000-0002-4928-4003]{Arjun~Dey}
\affiliation{NSF NOIRLab, 950 N. Cherry Ave., Tucson, AZ 85719, USA}

\author[0000-0002-2949-2155]{M.~Siudek}
\affiliation{Institute of Space Sciences, ICE-CSIC, Campus UAB, Carrer de Can Magrans s/n, 08913 Bellaterra, Barcelona, Spain}
\affiliation{Instituto de Astrof\'{\i}sica de Canarias, C/ V\'{\i}a L\'{a}ctea, s/n, E-38205 La Laguna, Tenerife, Spain}

\author[0000-0003-4440-259X]{M.~Mezcua}
\affiliation{Institut d'Estudis Espacials de Catalunya (IEEC), 08034 Barcelona, Spain}
\affiliation{Institute of Space Sciences, ICE-CSIC, Campus UAB, Carrer de Can Magrans s/n, 08913 Bellaterra, Barcelona, Spain}

\author[0000-0002-2733-4559]{J.~Moustakas}
\affiliation{Department of Physics and Astronomy, Siena College, 515 Loudon Road, Loudonville, NY 12211, USA}

\author[0000-0001-5537-4710]{S.~BenZvi}
\affiliation{Department of Physics \& Astronomy, University of Rochester, 206 Bausch and Lomb Hall, Rochester, NY 14627-0171, USA}

\author[0000-0003-4565-8239]{K.~Hainline}
\affiliation{Steward Observatory, University of Arizona, 933 North Cherry Avenue, Tucson, AZ 85721, USA}

\author[0000-0002-4684-9005]{R.~Hviding}
\affiliation{Max-Planck-Institut f¨ur Astronomie, K¨onigstuhl 17, D-69117 Heidelberg, Germany}
\affiliation{Steward Observatory, University of Arizona, 933 North Cherry Avenue, Tucson, AZ 85721, USA}

\author[0000-0002-1200-0820]{Yao-Yuan Mao}
\affiliation{Department of Physics and Astronomy, University of Utah, 115 South 1400 East, Salt Lake City, UT 84112, USA}

\author[0000-0002-5896-6313]{D.~M.~Alexander}
\affiliation{Centre for Extragalactic Astronomy, Department of Physics, Durham University, South Road, Durham, DH1 3LE, UK}
\affiliation{Institute for Computational Cosmology, Department of Physics, Durham University, South Road, Durham DH1 3LE, UK}

\author{R.~Alfarsy}
\affiliation{Institute of Cosmology and Gravitation, University of Portsmouth, Dennis Sciama Building, Portsmouth, PO1 3FX, UK}

\author[0000-0001-8522-9434]{C.~Circosta}
\affiliation{European Space Agency (ESA), European Space Astronomy Centre (ESAC), Camino bajo del Castillo s/n, Villanueva de la Canada, E28692 Madrid, Spain}
\affiliation{Department of Physics \& Astronomy, University College London, Gower Street, London, WC1E 6BT, UK}

\author[0000-0001-9457-0589]{Wei-Jian Guo}
\affiliation{National Astronomical Observatories, Chinese Academy of Sciences, A20 Datun Rd., Chaoyang District, Beijing, 100012, P.R. China}

\author[0000-0002-7113-0262]{V.~Manwadkar}
\affiliation{Department of Physics, Stanford University, 382 Via Pueblo Mall, Stanford, CA 94305, USA}
\affiliation{Kavli Institute for Particle Astrophysics and Cosmology, SLAC National Accelerator Laboratory, 2575 Sand Hill Road, Menlo Park, CA 94025, USA}

\author[0000-0002-4279-4182]{P.~Martini}
\affiliation{Center for Cosmology and AstroParticle Physics, The Ohio State University, 191 West Woodruff Avenue, Columbus, OH 43210, USA}
\affiliation{Department of Astronomy, The Ohio State University, 4055 McPherson Laboratory, 140 W 18th Avenue, Columbus, OH 43210, USA}
\affiliation{The Ohio State University, Columbus, 43210 OH, USA}

\author{B.~A.~Weaver}
\affiliation{NSF NOIRLab, 950 N. Cherry Ave., Tucson, AZ 85719, USA}

\author{J.~Aguilar}
\affiliation{Lawrence Berkeley National Laboratory, 1 Cyclotron Road, Berkeley, CA 94720, USA}

\author[0000-0001-6098-7247]{S.~Ahlen}
\affiliation{Physics Dept., Boston University, 590 Commonwealth Avenue, Boston, MA 02215, USA}

\author[0000-0001-9712-0006]{D.~Bianchi}
\affiliation{Dipartimento di Fisica ``Aldo Pontremoli'', Universit\`a degli Studi di Milano, Via Celoria 16, I-20133 Milano, Italy}

\author{D.~Brooks}
\affiliation{Department of Physics \& Astronomy, University College London, Gower Street, London, WC1E 6BT, UK}

\author{R.~Canning}
\affiliation{Institute of Cosmology and Gravitation, University of Portsmouth, Dennis Sciama Building, Portsmouth, PO1 3FX, UK}

\author{T.~Claybaugh}
\affiliation{Lawrence Berkeley National Laboratory, 1 Cyclotron Road, Berkeley, CA 94720, USA}

\author{K.~Dawson}
\affiliation{Department of Physics and Astronomy, The University of Utah, 115 South 1400 East, Salt Lake City, UT 84112, USA}

\author[0000-0002-1769-1640]{A.~de la Macorra}
\affiliation{Instituto de F\'{\i}sica, Universidad Nacional Aut\'{o}noma de M\'{e}xico,  Cd. de M\'{e}xico  C.P. 04510,  M\'{e}xico}

\author[0000-0002-5665-7912]{Biprateep~Dey}
\affiliation{Department of Astronomy \& Astrophysics, University of Toronto, Toronto, ON M5S 3H4, Canada}
\affiliation{Department of Physics \& Astronomy and Pittsburgh Particle Physics, Astrophysics, and Cosmology Center (PITT PACC), University of Pittsburgh, 3941 O'Hara Street, Pittsburgh, PA 15260, USA}

\author{P.~Doel}
\affiliation{Department of Physics \& Astronomy, University College London, Gower Street, London, WC1E 6BT, UK}

\author[0000-0002-3033-7312]{A.~Font-Ribera}
\affiliation{Department of Physics \& Astronomy, University College London, Gower Street, London, WC1E 6BT, UK}
\affiliation{Institut de F\'{i}sica d’Altes Energies (IFAE), The Barcelona Institute of Science and Technology, Campus UAB, 08193 Bellaterra Barcelona, Spain}

\author[0000-0002-2890-3725]{J.~E.~Forero-Romero}
\affiliation{Departamento de F\'isica, Universidad de los Andes, Cra. 1 No. 18A-10, Edificio Ip, CP 111711, Bogot\'a, Colombia}
\affiliation{Observatorio Astron\'omico, Universidad de los Andes, Cra. 1 No. 18A-10, Edificio H, CP 111711 Bogot\'a, Colombia}

\author{E.~Gaztañaga}
\affiliation{Institut d'Estudis Espacials de Catalunya (IEEC), 08034 Barcelona, Spain}
\affiliation{Institute of Cosmology and Gravitation, University of Portsmouth, Dennis Sciama Building, Portsmouth, PO1 3FX, UK}
\affiliation{Institute of Space Sciences, ICE-CSIC, Campus UAB, Carrer de Can Magrans s/n, 08913 Bellaterra, Barcelona, Spain}

\author[0000-0003-3142-233X]{S.~Gontcho A Gontcho}
\affiliation{Lawrence Berkeley National Laboratory, 1 Cyclotron Road, Berkeley, CA 94720, USA}

\author{G.~Gutierrez}
\affiliation{Fermi National Accelerator Laboratory, PO Box 500, Batavia, IL 60510, USA}

\author{K.~Honscheid}
\affiliation{Center for Cosmology and AstroParticle Physics, The Ohio State University, 191 West Woodruff Avenue, Columbus, OH 43210, USA}
\affiliation{Department of Physics, The Ohio State University, 191 West Woodruff Avenue, Columbus, OH 43210, USA}
\affiliation{The Ohio State University, Columbus, 43210 OH, USA}

\author{R.~Kehoe}
\affiliation{Department of Physics, Southern Methodist University, 3215 Daniel Avenue, Dallas, TX 75275, USA}

\author[0000-0003-2644-135X]{S.~E.~Koposov}
\affiliation{Institute for Astronomy, University of Edinburgh, Royal Observatory, Blackford Hill, Edinburgh EH9 3HJ, UK}
\affiliation{Institute of Astronomy, University of Cambridge, Madingley Road, Cambridge CB3 0HA, UK}

\author{A.~Lambert}
\affiliation{Lawrence Berkeley National Laboratory, 1 Cyclotron Road, Berkeley, CA 94720, USA}

\author[0000-0003-1838-8528]{M.~Landriau}
\affiliation{Lawrence Berkeley National Laboratory, 1 Cyclotron Road, Berkeley, CA 94720, USA}

\author[0000-0001-7178-8868]{L.~Le~Guillou}
\affiliation{Sorbonne Universit\'{e}, CNRS/IN2P3, Laboratoire de Physique Nucl\'{e}aire et de Hautes Energies (LPNHE), FR-75005 Paris, France}

\author[0000-0002-1125-7384]{A.~Meisner}
\affiliation{NSF NOIRLab, 950 N. Cherry Ave., Tucson, AZ 85719, USA}

\author{R.~Miquel}
\affiliation{Instituci\'{o} Catalana de Recerca i Estudis Avan\c{c}ats, Passeig de Llu\'{\i}s Companys, 23, 08010 Barcelona, Spain}
\affiliation{Institut de F\'{i}sica d’Altes Energies (IFAE), The Barcelona Institute of Science and Technology, Campus UAB, 08193 Bellaterra Barcelona, Spain}

\author[0000-0001-7145-8674]{F.~Prada}
\affiliation{Instituto de Astrof\'{i}sica de Andaluc\'{i}a (CSIC), Glorieta de la Astronom\'{i}a, s/n, E-18008 Granada, Spain}

\author{G.~Rossi}
\affiliation{Department of Physics and Astronomy, Sejong University, Seoul, 143-747, Korea}

\author[0000-0002-9646-8198]{E.~Sanchez}
\affiliation{CIEMAT, Avenida Complutense 40, E-28040 Madrid, Spain}

\author{D.~Schlegel}
\affiliation{Lawrence Berkeley National Laboratory, 1 Cyclotron Road, Berkeley, CA 94720, USA}

\author{M.~Schubnell}
\affiliation{Department of Physics, University of Michigan, Ann Arbor, MI 48109, USA}
\affiliation{University of Michigan, Ann Arbor, MI 48109, USA}

\author[0000-0002-6588-3508]{H.~Seo}
\affiliation{Department of Physics \& Astronomy, Ohio University, Athens, OH 45701, USA}

\author{D.~Sprayberry}
\affiliation{NSF NOIRLab, 950 N. Cherry Ave., Tucson, AZ 85719, USA}

\author[0000-0003-1704-0781]{G.~Tarl\'{e}}
\affiliation{University of Michigan, Ann Arbor, MI 48109, USA}

\author[0000-0002-6684-3997]{H.~Zou}
\affiliation{National Astronomical Observatories, Chinese Academy of Sciences, A20 Datun Rd., Chaoyang District, Beijing, 100012, P.R. China}





\begin{abstract}

Using early data from the Dark Energy Spectroscopic Instrument (DESI) survey, we search for AGN signatures in 410,757 line-emitting galaxies. By employing the BPT emission-line ratio diagnostic diagram, we identify AGN in 75,928/296,261 ($\approx$25.6\%) high-mass ($\logmass >$ 9.5) and 2,444/114,496 ($\approx$2.1\%) dwarf ($\logmass \leq$ 9.5) galaxies. Of these AGN candidates, 4,181 sources exhibit a broad $\ha$ component, allowing us to estimate their BH masses via virial techniques. This study more than triples the census of dwarf AGN and doubles the number of intermediate-mass black hole (IMBH;~$\mbh \le 10^6~\msun$) candidates, spanning a broad discovery space in stellar mass (7~$< \logmass <$~12) and redshift (0.001 $< \rm z <$ 0.45). The observed AGN fraction in dwarf galaxies ($\approx$2.1\%) is nearly four times higher than prior estimates, primarily due to DESI's smaller fiber size, which enables the detection of lower luminosity dwarf AGN candidates. We also extend the $\mbh - \mstar$ scaling relation down to $\logmass \approx$ 8.5 and $\logmbh \approx$ 4.4, with our results aligning well with previous low-redshift studies. The large statistical sample of dwarf AGN candidates from current and future DESI releases will be invaluable for enhancing our understanding of galaxy evolution at the low-mass end of the galaxy mass function. 

\end{abstract}



\section{Introduction} \label{sec:intro}

Supermassive Black Holes (SMBHs; $\mbh \gtrsim 10^{6}~\msun$) are found at the centers of all bulge-dominated galaxies with stellar masses $\mstar \gtrsim 10^{10}~\msun$ \citep{Kormendy&Richstone1995, Magorrian+1998, Kormendy&Ho2013}, yet their origin is far from understood. Theoretical studies suggest that they grow via accretion and mergers from initial ``seed'' black holes (BHs) formed early in the Universe \citep{Volonteri2010}. While these early BHs at high redshifts are difficult to detect with current capabilities, dwarf ($\mstar \le 3 \times 10^{9}~\msun$) galaxies in the nearby universe can host lower mass central BHs that can help constrain the BH seed formation models \citep{Volonteri2010, Greene+2012}. 

The search for central BHs in dwarf galaxies, especially for the still elusive Intermediate-mass Black Holes (IMBHs; $\mbh \le 10^{6}~\msun$), has been actively pursued over the past decade \citep[see][for reviews]{Mezcua2017, Greene+2020, Reines2022}. The accretion of matter onto these BHs powers active galactic nuclei (AGN), releasing a large amount of energy across the electromagnetic spectrum. In the optical regime, emission lines arising from ionized gas can be used to search for signatures of AGN photoionization. For instance, the BPT \citep[``Baldwin, Phillips \& Terlevich'';][]{bpt, bpt_vo87} emission-line diagnostic diagrams have been widely employed to identify hundreds of dwarf AGN candidates \citep{Reines+2013, Moran+2014, Mezcua+2020, Molina+2021, Polimera+2022, Salehirad+2022, Siudek+2023, Mezcua+2024a}. Beyond optical diagnostics, other methods such as infrared color-color diagrams \citep{Kaviraj+2019, Lupi+2020, Latimer+2021a}, radio observations \citep{Mezcua+2019, Reines+2020, Davis+2022}, X-ray emission \citep{Lemons+2015, Pardo+2016, Mezcua+2016,  Mezcua+2018, Birchall+2020, Latimer+2021b, Bykov+2024, Sacchi+2024}, and variability techniques \citep{Baldassare+2020, Burke+2022, Ward+2022} have been utilized to increase the census of dwarf AGN candidates. Despite this progress, questions remain regarding the prevalence of AGN in dwarf galaxies. 

Optical spectroscopy also offers an avenue to infer key physical properties of AGN, including their luminosity and BH masses. The gravitational influence of the BH accelerates the gas in its vicinity, resulting in broad emission lines in the spectrum. The broad $\ha$ emission from single-epoch spectroscopy is frequently used to estimate BH masses in galaxies through virial techniques \citep{Greene&Ho2005, Reines+2013, Moran+2014, Baldassare+2015, Reines&Volonteri2015, Chilingarian+2018, Suh+2020}. \citet{Baldassare+2015} identified a BH with a mass of $\approx$50,000 $\msun$ in a nearby dwarf galaxy, which is the lowest mass BH detected in a galaxy so far. However, further efforts to identify more of such lower-mass BHs are essential for discerning the BH seed formation models. 

Scaling relations between mass of the central SMBH and various galaxy properties such as the mass/luminosity of the stellar bulge \citep{Ferrarese&Merritt2000, Marconi&Hunt2003, Lauer+2007, Gultekin+2009, McConnell&Ma2013}, the velocity dispersion of stars in the bulge \citep{Ferrarese&Merritt2000, Gebhardt+2000, Gultekin+2009, McConnell&Ma2013}, and the galaxy stellar mass \citep{Reines&Volonteri2015, Suh+2020}, have been extensively studied. The tightness of the observed correlations suggests that SMBHs and their host galaxies co-evolve \citep{Kauffmann&Haehnelt2000, Kormendy&Ho2013}. However, whether this co-evolution extends to lower-mass galaxies is unknown. 

The low-mass end of the scaling relations can provide insight into the BH seed formation mechanisms and the efficiency of BH growth in dwarf galaxies \citep{Greene+2020}. By measuring the stellar velocity dispersion ($\sigma_{\star}$) of eight broad-line candidates from \citet{Reines+2013}, \citet{Baldassare+2020} extended the $\mbh - \sigma_{\star}$ relation down to \logmass~$\approx$ 8.96. They found that the dwarf galaxies are in good agreement with the extrapolation of the existing relation \citep{Kormendy&Ho2013}. In contrast, using a sample of 127 low-mass Seyfert 1 galaxies, \citet{Martin-Navarro&Mezcua2018} found a flattening at the low-mass end of the $\mbh - \sigma_{\star}$ relation. Measuring $\sigma_{\star}$ in low-mass systems is difficult to extend to a larger sample and to farther distances due to the faintness of targets and the limitations of current telescopes and instrumentation. Therefore, even though the $\mbh - \sigma_{\star}$ relation usually shows the tightest correlation, the $\mbh - \mstar$ relation is preferred for both extending the scaling relations to lower galaxy masses and/or for comparing to higher redshifts. 

Using a sample of 262 broad-line AGN and 79 galaxies with dynamical BH masses, including fourteen dwarf galaxies, \citet{Reines&Volonteri2015} measured the $\mbh - \mstar$ relation down to $\logmass \approx$ 9.0. They found two distinct scaling relations for early-type and late-type galaxies, with the former having a higher normalization than the latter and the dwarf galaxies falling on the tail of the late-type galaxy relation. More recently, \citet{Mezcua+2023} and \citet{Mezcua+2024b} found nineteen low-mass galaxies at z $\rm\approx 0.9 - 3$ with over-massive BHs that reside $\gtrsim$ 2 dex above the $\mbh - \mstar$ relation. Several such over-massive BHs were also discovered at higher redshifts, z $\rm\approx 4 - 11$ using observations from the James Webb Space Telescope \citep[JWST;][]{Harikane+2023, Maiolino+2023, Ubler+2023}. However, these high-redshift studies are subject to observational bias, as they predominantly detect luminous AGN. Overall, these studies are constrained by limited sample sizes at the low-mass end, and the scaling relation remains poorly constrained at both low stellar and BH masses, even at low redshifts. 

In this paper, we use early spectroscopic data from the Dark Energy Spectroscopic Instrument \citep[DESI;][]{desi_exp1, desi_exp2, desi_instrument_overview} survey to identify AGN in galaxies spanning a wide range of stellar masses ($6 \le \logmass < 12$), specifically to increase the census of AGN in dwarf galaxies, and to extend the $\mbh - \mstar$ scaling relation down to lower galaxy masses. The early DESI data is the largest sample of optical spectra to date, encompassing nearly 1.5 million unique galaxies at z $\le$ 0.45 for our analysis. 

This paper is organized as follows. Section~\ref{sec:data} describes the spectroscopic and photometric data used in the paper. We present the estimation of stellar masses and emission-line measurements of the targets in Section~\ref{sec:methods}. Sections~\ref{sec:agn_selection} and \ref{sec:scaling_relation} details the AGN identification and the overall $\mbh - \mstar$ scaling relation, respectively. We discuss our results in Section~\ref{sec:discussion} and summarize our conclusions in Section~\ref{sec:conclusions}. Throughout the paper, we assume the Chabrier initial mass function \citep[IMF; ][]{chabrier_imf}, and the \citet{planck_cosmology} cosmology with $H_{0}$ = 67.4 $\rm km~s^{-1}~Mpc^{-1}$ and $\Omega$ = 0.315. All wavelengths are presented in vacuum wavelengths, and all magnitudes are given in the AB system \citep{AB_Mag_System}.

\section{Data} \label{sec:data}
\subsection{Spectroscopy} \label{subsec:spectroscopy}
DESI is a 5000-fiber multi-object spectrograph on the 4-meter Mayall telescope at Kitt Peak National Observatory (KPNO), covering a spectral range of 3600 \AA~to 9800 \AA. The resolving power of the instrument ranges from 2000 at 3600 \AA~to 5500 at 9800 \AA~\citep{desi_instrument_overview, desi_focal_plane, desi_corrector}. This instrument will be used to conduct a cosmological survey over an area of $\approx$14,000 $\rm deg^{2}$ over five years and will obtain optical spectra of nearly 35 million galaxies and quasars \citep{desi_exp, desi_exp1, desi_exp2, desi_ops}. DESI completed its survey validation (SV) observations in May 2021, and is currently carrying out its main survey. The SV data, covering $\approx$1,400 $\rm deg^{2}$ of the sky, is released as the DESI Early Data Release \citep[EDR;][]{desi_edr2, desi_edr1}. We use the entire DESI EDR and 20\% of Year 1 (DA0.2) data for analysis in this paper.

DESI targets are divided into five primary classes: 1) Milky Way Survey \citep[{\tt MWS};][]{mws}, 2) Bright Galaxy Survey \citep[{\tt BGS};][]{bgs1, bgs2, Juneau+2024}, 3) Luminous Red Galaxies \citep[\tt{LRG};][]{lrg1, lrg2}, 4) Emission Line Galaxies \citep[{\tt ELG};][]{elg1, elg2}, and 5) Quasars \citep[{\tt QSO};][]{qso1, qso2}. These are further complemented by secondary targets via spare fibers \citep[{\tt SCND}; see Appendix of][]{desi_edr2, Darragh-Ford+2023, Fawcett+2023}. This target selection is validated based on a thorough visual inspection of the SV spectra \citep{Alexander+2023, Lan+2023}. Depending on the selection criteria, some sources can be targeted in multiple targeting classes. The DESI targeting selection is described in detail by \citet{desi_tgts}. The biases related to this target selection on our measurements are briefly discussed in Sections~\ref{subsec:mstar} and \ref{subsec:agn_frac}.

The spectra obtained from DESI are reduced using the DESI spectroscopic pipeline \citep{desi_spec_pipeline}, and the redshifts are obtained by the {\tt Redrock} redshift-fitting pipeline \citep[][Bailey et al., in preparation]{redrock_qso, Anand+2024}. The DESI EDR and DA0.2 spectra are reduced and released as {\it `fuji'} and {\it `guadalupe'} internal data releases, respectively. Combining the redshift catalogs from both these releases results in 6,059,937 spectra of 5,859,563 unique targets. For objects with multiple spectra, we choose the ``best spectrum'' based on the {\tt ZCAT\_PRIMARY} flag. This prioritizes the spectrum that has no issues with fiber and redshift measurements. If multiple spectra pass these criteria, then we choose the highest signal-to-noise ratio spectrum as the ``best spectrum'' for the object (see Section 3.3.3 in \citet{desi_edr2} for more information). After selecting the unique spectra of all objects, we apply the following data quality cuts: {\tt COADD\_FIBERSTATUS = 0}\footnote{\url{https://desidatamodel.readthedocs.io/en/latest/bitmasks.html\#fiberstatus-bit-definitions}}, {\tt ZWARN = 0 or 4}\footnote{\url{https://desidatamodel.readthedocs.io/en/latest/bitmasks.html\#zwarn-bit-definitions}}. This selects 4,323,517 sources that have been observed without any fiber issues and have good redshift measurements. Of these, we select sources that are assigned spectroscopic types of {\tt GALAXY} or {\tt QSO} by {\tt Redrock}, resulting in a total of 3,483,455 sources. Among these, 1,628,394 sources have redshifts in the range of 0.001 $\le\rm z \le$ 0.45. The lower redshift limit is to exclude contamination from Galactic sources. Given that we focus on identifying AGN candidates using standard emission-line ratio diagnostics (Section~\ref{subsec:bpt_diagram}), the upper redshift limit is determined by the availability of \siilam~emission lines in the DESI spectra (see Section~\ref{subsec:emfit} for more information). 

For some {\tt QSO} targets, the redshifts determined by {\tt Redrock} were found to be incorrect, resulting in the presence of high-redshift contaminants in our sample \citep{Alexander+2023}. To accurately determine their redshifts, the DESI team used a machine learning algorithm called {\tt QuasarNet}, which is a deep convolutional neural network classifier specifically designed to identify quasars and their redshifts.  Of the 1,628,394 {\tt Redrock} redshifts, we find that 4,286 are classified as high-redshift ($\rm z \gtrsim$ 0.5 $-$ 4) quasars by {\tt QuasarNet}. We remove these from our selection, resulting in a spectroscopic sample of 1,624,108 sources in our redshift range of interest.   

\subsection{Photometry} \label{subsec:photometry}

The primary DESI targets are selected based on the DESI Legacy imaging surveys \citep{desi_imaging}. The ninth data release for this survey (DR9) comprises a set of optical and infrared imaging data covering the entire 14,000 $\rm deg^{2}$ of the DESI survey (Schlegel et al., in preparation). The photometric catalog consists of optical imaging data from three different surveys: The Beijing-Arizona Sky Survey \citep[BASS;][]{bass}, the DECam Legacy Survey (DECaLS), and the Mayall $z$-band Legacy Survey (M$z$LS). Furthermore, it also includes the WISE/NEOWISE \citep[unWISE; ][]{Lang+2014, Meisner+2016, Meisner+2017} coadd images from the Wide-field Infrared Space Explorer \citep[WISE;][]{wise}. 

Imaging from the Legacy Surveys is processed using inference modeling with the {\it Tractor \footnote{\url{https://github.com/dstndstn/tractor}}} code \citep{Lang+2016}. This produces a catalog that contains photometry in $g$, $r$, and $z$ bands, based on an optimal model of the source morphology. It also includes mid-IR photometry measured from the forced photometry of the unWISE coadds using the optically derived model. This detects fainter sources than the traditional approach of WISE photometry \citep{Lang+2016}. Furthermore, the morphology ({\tt TYPE}) and S{\'e}rsic index ({\tt SERSIC}) of each source are also computed based on this model.

The optical photometry is corrected for Galactic extinction using dust emission maps from \citet{extinction_sfd98}. The extinction values for the WISE filters are derived from \citet{extinction_fitzpatrick1998}, using the recommendations by \citet{extinction_sf2011}. These are reported in the LS DR9 catalog in linear units of Milky Way transmission. 

In addition to the primary DESI targets, the spectral data from the survey also includes secondary targets as well as targets of opportunity, which might not always have the Legacy Survey photometry. Given that our analysis requires the stellar mass calculation using this photometry, we select the 1,623,425 objects that have DR9 photometry. Additionally, we apply two quality cuts that remove false detections, extremely faint sources, and a significant fraction of fragmented sources: 

\begin{equation*}
    {\rm SNR \ge 5~in~}g,r,z~{\rm bands}
\end{equation*}
\begin{equation*}
    {\rm FRACFLUX \le 0.25~in~}g,r,z~{\rm bands}
\end{equation*}

We do not apply any cuts on the WISE photometry to include faint dwarf galaxies, which might not always have good available WISE detections. Given the strong cuts on optical photometry, we will instead focus on quality cuts based on stellar mass estimates (Section~\ref{subsec:mstar}). The above cuts result in 1,492,821 unique sources with good spectra and valid photometry as the starting sample for our analysis.

\section{Methods} \label{sec:methods}
\subsection{Stellar Masses} \label{subsec:mstar}

\begin{figure*}
    \centering
    \includegraphics[width = \textwidth]{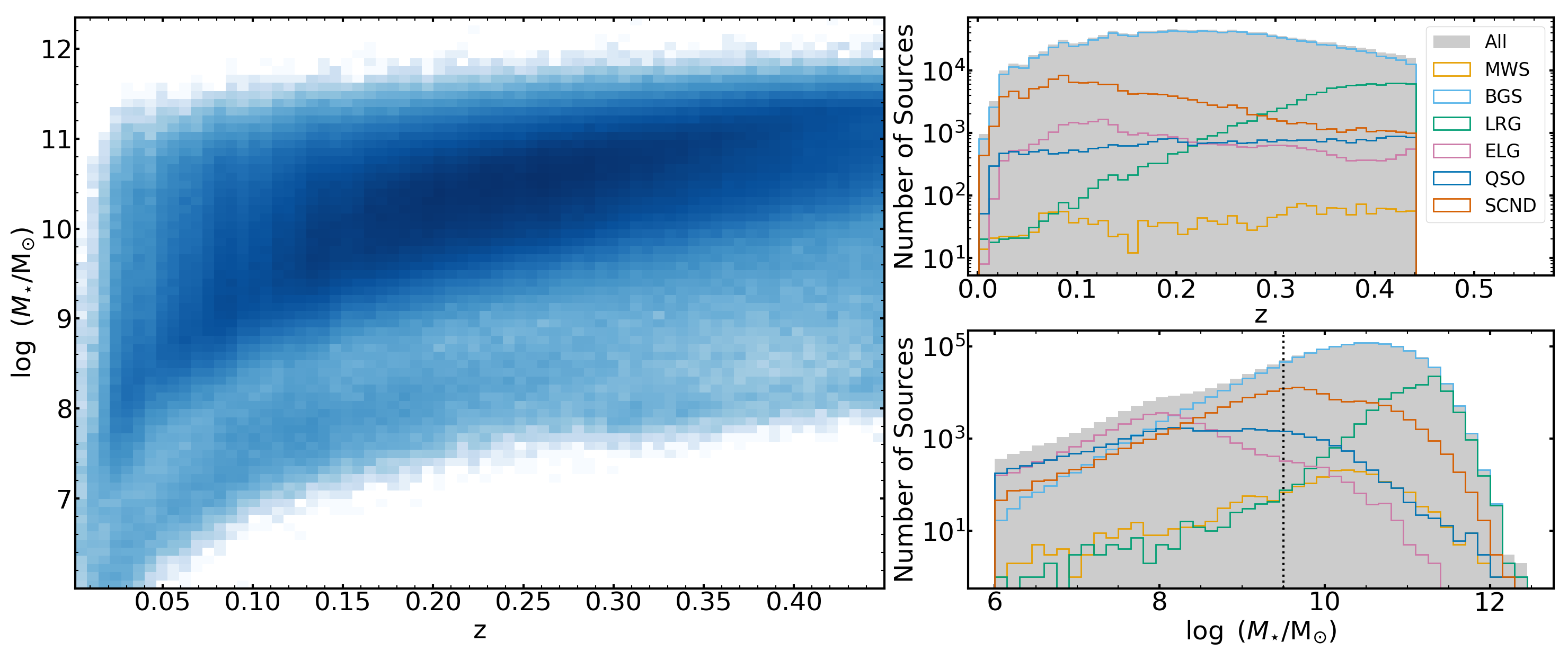}
    \caption{{\it Left:} 2D density distribution of 1,358,177 sources in the $\logmass - \rm z$ space. {\it Top Right:} Redshift distribution of the sources divided by their DESI targeting type. The underlying gray histogram is the redshift distribution of all the sources. {\it Bottom Right: } Stellar mass distribution of the sources divided by their DESI targeting type. The underlying gray histogram is the stellar mass distribution of all the sources. The vertical dotted line denotes the separation of dwarf ($\logmass \leq$ 9.5) galaxies and high-mass ($\logmass >$ 9.5) galaxies. Some objects are targeted in more than one target class, therefore, some sources are duplicated in different histograms. The stellar mass and redshift distributions of the galaxies are influenced by DESI target selection, noting that the majority of galaxies in our sample are from {\tt BGS}.} 
    \label{fig:mstar_dist}
\end{figure*}

We fit the spectral energy distributions (SED) using Code Investigating GALaxy Emission \citep[CIGALE v.22.1;][]{cigale} to estimate the stellar masses of our sample while accounting for AGN contribution. We use \citet{bc03} single stellar population models assuming a Chabrier IMF \citep{chabrier_imf} and solar metallicity. Furthermore, we consider a delayed star formation history (SFH) model with an optional exponential burst to generate the grid of models. We include the standard nebular emission model from \citet{Inoue2011} and the dust attenuation model using the \citet{calzetti_dust} attenuation curve. The reprocessed dust emission is modeled adopting the dust emission models of \citet{Dale+2014}, whereas the AGN emission is modeled using the models from \citet{Fritz+2006}. CIGALE simultaneously fits the AGN and the galaxy component using all available photometry ($g$, $r$, $z$, W1, W2, W3, and W4) at the spectroscopic redshift of each galaxy. It returns the estimates for the main galaxy properties, including stellar mass, star formation rate, and the relative contribution of the dusty AGN torus to the total infrared luminosity. In cases where the fit is consistent with no or negligible AGN contribution, CIGALE outputs the AGN fraction to be zero. The quality of fit is expressed by reduced $\chi^2$, and the estimates and errors of physical properties are the likelihood-weighted mean and standard deviations of the probability distribution function, respectively \citep{cigale}. The physical properties of galaxies and AGN for both {\it fuji} and {\it guadalupe} releases are generated as value-added catalogs (VACs), and the former is now publicly available with DESI EDR\footnote{\url{https://data.desi.lbl.gov/doc/releases/edr/vac/cigale}} \citep[][]{Siudek+2024}.

CIGALE performs better when mid-IR photometry is available, but it can still converge to a reliable solution using confident optical photometry (in $grz$ bands) along with any available WISE photometry \citep{Siudek+2024}. We therefore do not place any constraints on the WISE photometry of sources (Section~\ref{subsec:photometry}), and instead focus on the following quality cuts to select sources with confident stellar masses:

\begin{equation*}
    \chi_{\rm CIGALE}^{2} \le 10
\end{equation*}
\begin{equation*}
    \logmass \ge 6
\end{equation*}
\begin{equation*}
    {\rm Error~in}~\logmass \le 0.5~{\rm dex}
\end{equation*}

CIGALE computes two estimates of the different parameters based on the best-fit model ({\tt best}) and the likelihood-weighted mean measured from the probability density function marginalized over all the parameters ({\tt bayes}). The comparison between the two measurements is parametrized by {\tt FLAG\_MASSPDF}, which is defined as $\log M_{\rm\tt best}/\log M_{\rm\tt bayes}$. To select accurate stellar mass measurements, we further apply the following cut as recommended by the CIGALE VAC \citep[see][]{Mountrichas+2021, Siudek+2024}.

\begin{equation*}
    0.2~\le~{\tt FLAG\_MASSPDF}~\le~5
\end{equation*}

This results in 1,358,177 sources with stellar masses ranging from \logmass~= 6.0 $-$ 12.3, with a median of \logmass~= 10.3. For consistency with similar studies on dwarf AGN candidates, we define galaxies with $\logmass \le 9.5$ as dwarf galaxies throughout this paper. This threshold is approximately equivalent to the stellar mass of LMC \citep{van_der_Marel+2002}, the largest satellite galaxy of the Milky Way.

The left panel of Figure~\ref{fig:mstar_dist} shows the 2D distribution of all 1,358,177 sources in the $\logmass - \rm z$ space. The source density peaks at $\logmass \approx$ 10.5 and $\rm z \approx$ 0.2. We find that the stellar masses at $\rm z \approx 0.45$ extend down to $\logmass \approx 7.5$. There are two overlapping distributions varying with redshift, with the higher-density sample extending from low-redshift dwarf galaxies to high-redshift massive galaxies. The lower-density sample hosts the majority of dwarf galaxies in our sample. This is primarily due to the DESI targeting selection (see Section~\ref{subsec:spectroscopy}) as shown on the right panel of the Figure. The top-right panel shows the redshift distribution of sources, divided by their DESI target type. All the target types are uniformly distributed across our redshift range of interest, except the {\tt LRG} targets that have a redshift distribution skewed toward the high end of the redshift range.  The {\tt MWS} sources are intended to be Milky Way stars by the targeting criteria, but {\tt Redrock} recovers them as {\tt GALAXY} or {\tt QSO}. We also see that {\tt BGS} targets dominate across all redshifts. This is expected because BGS is designed to focus on low redshifts, while {\tt LRG}, {\tt ELG}, and {\tt QSO} targets focus on $\rm z > 0.5$. The bottom-right panel shows the stellar mass distribution of the sources, distributed between different target types. The dotted vertical line denotes the separation of dwarf galaxies from high-mass galaxies that we consider in this paper. Of the 1,385,177 total sources, we have 222,680 dwarf galaxies (with $\logmass \le$ 9.5) and 1,135,497 high-mass galaxies (with $\logmass >$ 9.5). The {\tt BGS} targets constitute the majority of both dwarf galaxies and high-mass galaxies in our sample. The rest of the dwarf galaxies come from {\tt SCND}, {\tt ELG}, and {\tt QSO}, with $<$1\% sources from {\tt LRG} and {\tt MWS}. On the other hand, the remaining high-mass galaxies are primarily {\tt LRG} and {\tt SCND} targets and $<$1\% are from {\tt ELG}, {\tt QSO}, and {\tt MWS}. We discuss the impact of these targeting selections on AGN identification in Section~\ref{subsec:agn_frac}.

\subsection{Emission-Line Measurements} \label{subsec:emfit}

\begin{figure*}
    \centering
    \includegraphics[width=1.0\textwidth]{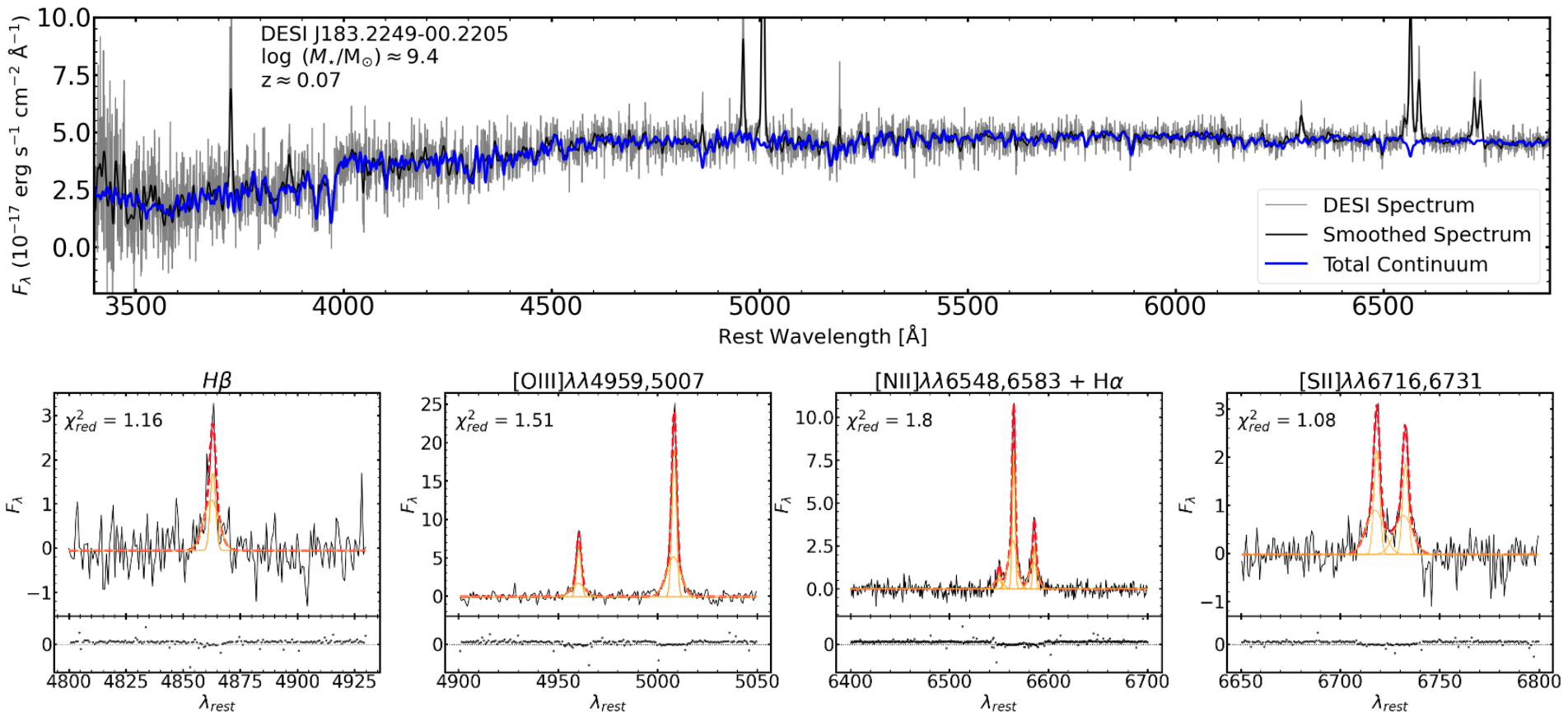}
    \caption{Example fit of a dwarf galaxy spectrum via the default mode. {\it Top:} The rest-frame spectrum, the smoothed spectrum, and the total continuum fit are plotted in gray, black, and blue, respectively. The stellar mass and redshift of the source are mentioned in the panel. {\it Bottom: } Best-fit models to the continuum-subtracted emission line spectrum in the regions of interest - $\hb$, \oiii, \nii + $\ha$, and \sii~- from left to right. The spectrum is shown in black, while the best-fit models are shown in dashed red. The individual narrow and outflow components are plotted in orange. The reduced $\chi^{2}$ values for each of the fits are given in the upper left corner of the panels. The fractional residuals are plotted as gray points in the bottom panels for each of the fits.}
    \label{fig:default-mode}
\end{figure*}

To accurately determine the emission-line ratios for AGN selection and to estimate the BH masses using virial techniques, we need careful measurements of the flux and width ($\sigma$) values of various emission lines. For this purpose, we developed a Python-based emission-line fitting code ({\tt EmFit}), which uses a non-linear least squares fitting algorithm, for fitting $\hb~\lambda$4861, \oiiilam, \niilamlam, $\ha~\lambda$6563, and \siilam~emission lines. The code includes testing for extra components for \oiii, \sii, and $\ha$. We run {\tt EmFit} on all 1,624,108 sources with good spectra, irrespective of photometry and stellar mass measurements. We describe the fitting procedure in this subsection. 

For each object, we correct the DESI spectrum for galactic reddening using the \citet{extinction_fitzpatrick1998} dust extinction law. We then obtain the stellar continuum from {\tt FastSpecFit}\footnote{\url{https://fastspecfit.readthedocs.io/en/latest}} \citep[][J.~Moustakas et al., in preparation]{fastspecfit}. We use {\tt FastSpecFit} v3.2 for {\it fuji} and v3.1 for {\it guadalupe}; the VAC for DESI-EDR is publicly available\footnote{\url{https://data.desi.lbl.gov/doc/releases/edr/vac/fastspecfit}}. {\tt FastSpecFit} models the stellar continuum using 168 composite stellar population templates of varying age, stellar metallicity, and dust attenuation. The spectra are then corrected for any residual, unmodeled flux by constructing a ``smooth'' continuum using a sliding median with iterative outlier-clipping. We subtract this total continuum from the observed spectrum and convert the resulting emission-line spectrum to rest-frame. 

We have two different fitting modes for the rest-frame spectra: the ``default'' mode and the ``Extreme Broad Line (EBL)'' mode. The EBL sources are defined here as sources where the broad component of $\ha$ extends up to the \sii~region. For selecting such sources, we compare the blue side of \sii~emission-lines (6600 $-$ 6670 \AA) to the red side (6700 $-$ 6900 \AA). If the difference between median flux densities from both sides exceeds $0.5 \times 10^{-17} \ergscmsq$, and if the flux density on the blue side is at least five times greater than the red side, we flag the source as an EBL candidate. Of the 1,624,108 sources, 3,457 sources (0.21\%) are fit via the EBL mode, while the rest are fit via the default mode.  

For both modes, we model each emission line using one or more Gaussian functions, which are constrained to have positive flux. We also include a zero-order polynomial to each fit, as a proxy for any residual continuum around the emission lines. Adding an extra Gaussian component to a fit typically improves the fit and decreases its reduced $\chi^{2}$. To evaluate the statistical significance of this improvement, we compute the $\chi^{2}$ difference, $\Delta\chi^{2}$, with and without the extra component. By Wilk's theorem, the $\Delta\chi^{2}$ asymptotically approaches a $\chi^{2}$ distribution under the null hypothesis of a simple fit with a single Gaussian \citep{Wilks1938}. Thus, we convert the $\Delta\chi^{2}$ into a tail probability ($p$-value), and we reject the null hypothesis when $p$-value is less than $\rm3\times10^{-7}$. This provides at least a 5$\sigma$ confidence for the existence of an extra component.

We first describe our default fitting method, followed by the EBL fitting method for the flagged sources. 

\subsubsection{Default Fitting Mode} \label{subsubsec:default_fit}
For the sources that are not flagged as EBL sources, we divide the rest-frame emission-line spectrum into four windows: 1) 4700~\AA~$\le~\lambda_{rest}~\le$ 4930~\AA~for the $\hb$ region; 2) 4900~\AA~ $\le~\lambda_{rest}~\le$ 5100~\AA~ for the \oiii~region; 3) 6300~\AA~ $\le~\lambda_{rest}~\le$ 6700~\AA~ for the \nii + $\ha$ region; and 4) 6630~\AA~ $\le~\lambda_{rest}~\le$ 6900~\AA~  for the \sii~region. 

We first fit the \sii~doublet with a pair of Gaussian profiles tied to have the same redshift and same intrinsic linewidth in velocity space. In addition to the one-component fit, we also test for a two-component fit to the \sii~doublet, such that the amplitude ratio of the two components is equal for the lines \citep[following][]{Reines+2013}. When the spectrum is noisy, there is a risk that the model includes a low-amplitude broad component to fit the continuum. To avoid this unphysical scenario, the width of the second component is allowed to vary up to 600 $\rm km~s^{-1}$. However, if in such a scenario, a single component has a width $\rm > 450 km~s^{-1}$, we revert back to a two-component model. We apply quality cuts based on this scenario to remove unphysical fits (see Section~\ref{subsubsec:quality_cuts}). The two-component model is accepted if the $p$-value from the $\Delta\chi^{2}$ provides a $>5\sigma$ confidence. In such a case, we label the narrower component as the primary component and the broader component as the secondary component. The second component can either be an ``outflow'' component or a second narrow peak in double-peaked emission line sources (see Appendix~\ref{app:double-peak}). Only 11,000 ($\approx$0.8\%) sources require a second component. The \sii~emission line profile has been shown to generally match other narrow emission lines \citep{Greene+2004, Reines+2013}. We, therefore, use the final \sii~model as a template for the \nii~and $\ha$ emission lines.

We then fit the \nii + $\ha$ complex, where the redshifts of the \nii~and $\ha$ lines are tied to the \sii~lines. The amplitude ratio of \nii$\lambda$6583/\nii$\lambda$6548 is fixed to the theoretical value of 2.96 \citep{Acker+1989}. When \sii~is best-fit with a single component model, the width of \nii~lines is fixed to the width of \sii~lines in velocity space, and the width of the narrow $\ha$ component is initially allowed to increase up to 30\% of the \sii~width \citep[similar to][]{Reines+2013}. If $\ha$ does not satisfy this constraint, the width of the narrow $\ha$ is also held fixed to that of the \sii~component. For the two-component \sii~model, all the three emission lines - \nii~doublet and $\ha$ - are also considered to have two components such that the width of the respective components is fixed to the \sii~components in velocity space.  Additionally, the separation between the two components is also fixed as measured from the \sii~template. Having the same second component as \sii~lines is important to achieve a physically probable broad component after removing the contribution from the second component (see Figure~\ref{fig:default-mode} and Appendix~\ref{app:vi}). For both one and two-component fits, we further test for a broad component ($\rm\ha;b$) with a minimum FWHM ($\rm\ha;b$) $\rm > 300~km~s^{-1}$, based on the $p$-value from the $\Delta\chi^{2}$ with and without a broad component. The broad component is accepted when it delivers $>5\sigma$ confidence. However, this component is rejected if this component is narrower than the second component. 

We use the resulting $\ha$ fit as a template for the $\hb$ fit and the number of components of $\hb$ is fixed to be the same as $\ha$. The redshifts and widths of all the available $\hb$ components (primary, second, and broad) are fixed to the widths of the respective $\ha$ components in velocity space. Only the amplitude of the components is allowed to vary. 

Given that the \oiii~profile can exhibit a blue-shifted outflow component and does not match the profile of other narrow emission lines, we fit the \oiii~doublet separately. We fix the relative redshifts of the peaks, fix the amplitude ratio of \oiii$\lambda$5007/\oiii$\lambda$4959 to 2.98 \citep{Dimitrijevic+2007}, and add the condition that the widths of the two lines are equal in velocity space. Both one and two-component models are tested and the two-component fit is accepted if the $p$-value from the $\Delta\chi^{2}$ satisfies the 5$\sigma$ confidence cut. Similar to the \sii~model, the narrower component is taken as the primary component, and the broader component can either be an outflow component or a second component (see Appendix~\ref{app:double-peak}).

Figure~\ref{fig:default-mode} shows an example of a dwarf galaxy ($\logmass \approx 9.4$) spectrum fit via the default mode. The top panel shows the best-fit of the stellar continuum (in blue) from {\tt FastSpecFit} overplotted on the DESI spectrum (in gray). The resulting emission line spectrum (in black) and the best-fit models (in red) are shown in the bottom panel. We see that all the emission lines contain a second component (a possible outflow in this case). The primary and outflow Gaussian components are shown in orange, and the reduced $\chi^{2}$ values are shown in each of the panels. This galaxy is a dwarf AGN candidate (see Section~\ref{sec:agn_selection}) and shows the importance of adding a second component in the narrow emission lines. 

\begin{figure*}
    \centering
    \includegraphics[width=1.0\textwidth]{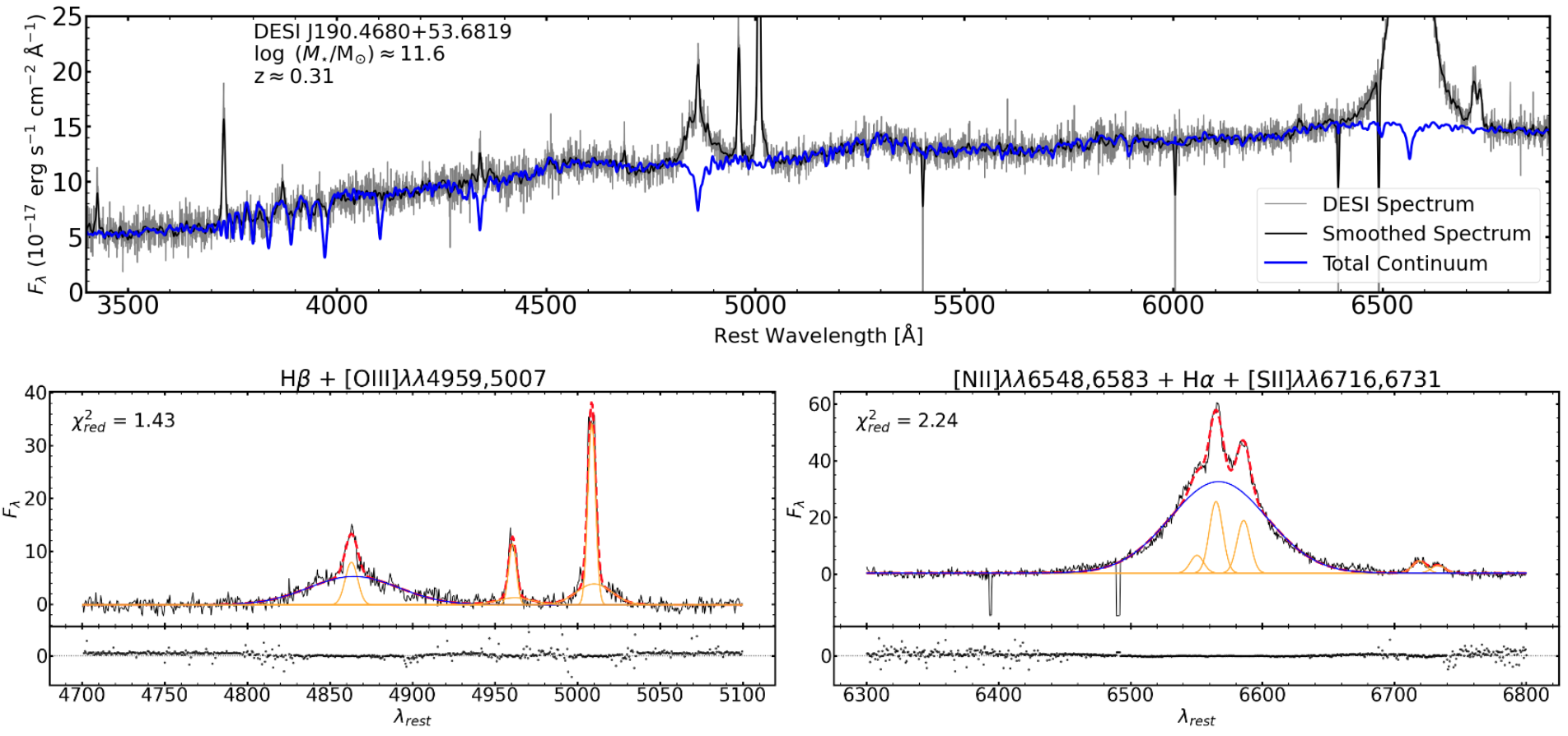}
    \caption{Example fit of a high-mass galaxy spectrum via the EBL mode. {\it Top:} The rest-frame spectrum, the smoothed spectrum, and the total continuum fit are plotted in gray, black, and blue, respectively. The stellar mass and redshift of the source are mentioned in the panel. {\it Bottom: } Best-fit models to the continuum-subtracted emission line spectrum in the regions of interest - $\hb$ + \oiii~(left) and \nii + $\ha$ + \sii~(right). The spectrum is shown in black, while the best-fit models are shown in dashed red. The individual narrow and outflow components are plotted in orange, while the broad Balmer components are plotted in blue. The reduced $\chi^{2}$ values for each of the fits are given in the upper left corner of the panels. The fractional residuals are plotted as gray points in the bottom panels for each of the fits.}
    \label{fig:ebl-mode}
\end{figure*}

\subsubsection{Extreme Broad Line Fitting Mode} \label{subsubsec:ebl_fit}
For the sources that are flagged as EBL sources, we divide the rest-frame emission-line spectrum into two windows: 1) 4700~\AA~ $\le~\lambda_{rest}~\le$ 5100~\AA~ for the $\hb$ + \oiii~region; and 2) 6300~\AA~ $\le~\lambda_{rest}~\le$ 6900~\AA~ for the \nii + $\ha$ + \sii~region. 

We first fit the \nii~ + $\ha$ + \sii~emission-lines with a single Gaussian component for each of the narrow lines and an extra broad Gaussian component. In the case of EBL sources, multiple broad components might be needed for properly fitting the complex emission line profile \citep{Greene&Ho2005, Liu+2019}. However, we approximate the complex spectrum with a single broad Gaussian component for this work. The amplitude ratio of \nii$\lambda$6583/\nii$\lambda$6548 is again fixed to the theoretical value of 2.96 \citep{Acker+1989}. The redshifts of all the narrow components (\nii~doublet, $\ha$, and \sii~doublet) are tied together and their widths are constrained to be equal in velocity space. In some sources, a noisy spectrum is selected as an EBL source. Their fits using this mode lead to an unrealistic broad $\ha$ component. Using visual inspection, we find that such sources are poorly fit with broad $\ha$ widths $\rm<~1000~km~s^{-1}$. We therefore use this criterion ($\rm\sigma (H\alpha;b) < 1000~km~s^{-1}$)  to pick such sources and rerun their fitting using the default mode. The median FWHM of broad $\ha$ components for the sources that are fit with the EBL mode is $\approx$2000~$\rm km~s^{-1}$.

For the $\hb$ + \oiii~region, we use the $\ha$ model as a template for the $\hb$ model and the \oiii~lines are fit independently. We fix the redshifts and widths of the $\hb$ components to scale with the $\ha$ components. The redshifts, amplitudes, and widths of the \oiii~components are tied as mentioned in the default mode. Both one- and two-component models are tested for the \oiii~lines, and the two-component fit is allowed if the $p$-value from the $\Delta\chi^{2}$ is less than the 5$\sigma$-confidence cut. 

Figure~\ref{fig:ebl-mode} shows an example of the best-fit of a high-mass EBL source ($\logmass \approx 11.6$). The panel shows the total continuum from {\tt FastSpecFit} in blue, overplotted on the spectrum. The bottom panels show the best-fits in the $\hb$ + \oiii~and \nii~+ $\ha$ + \sii~regions. The narrow and outflow components are shown in orange, while the broad Balmer components are shown in blue. The \oiii~lines in this galaxy clearly show evidence of outflows. 

\subsubsection{Measurements and Uncertainties} \label{subsubsec:measurements}

We compile the amplitude, mean, and standard deviation of all the detected Gaussian components from the best-fit model. The fluxes and widths ($\sigma$) are computed using these parameters and the resulting width is corrected for instrumental resolution by subtracting the resolution element in quadrature. To estimate the fit uncertainties, we consider a Gaussian error distribution with standard deviation derived from the inverse variance at each pixel of the spectrum. We vary the spectrum within this distribution and repeat the fitting procedure a hundred times. The uncertainties are estimated as the standard deviation of all measurements from these iterations. We assume that the pixels are uncorrelated and the errors are therefore underestimated. In addition to the flux and $\sigma$ measurements, we also measure the noise around the emission lines as the root-mean-square (rms) of the continuum around the lines. We also make note of {\tt PROB\_BROAD} as the fraction of iterations in which the broad $\ha$ component is measured with non-zero flux. This is especially important for selecting broad-line candidates when the spectra are noisy. 

The emission line ratio measurements require total fluxes from the narrow component of the emission lines (Section~\ref{sec:agn_selection}). As mentioned earlier, some of the second components in \sii,\oiii,\nii, and $\ha$ appear as narrow components for double-peaked emission line sources. We add the fluxes from the two components for such sources. The method of picking these double-peaked sources is described in Appendix~\ref{app:double-peak}.

\subsubsection{Applying Quality Cuts} \label{subsubsec:quality_cuts}
Even after the detailed careful fitting process, the fits to some noisy spectra lead to poor representations of the observations and unphysical flux and sigma values. One such instance is when the stellar continuum subtraction is erroneous and creates a non-physical jump between the continuum on either side of the emission lines. This is especially prominent in fitting \sii~lines, leading to broad second components. We remove these sources by applying the following cut: 

\begin{center}
    $\rm\sigma_{\sii,2} < 700~km~s^{-1}$  
\end{center}

\noindent where $\rm\sigma_{\sii,2}$ is the width of the second \sii~component. Another issue is related to the forced fitting of a negative continuum and an extremely high-velocity width narrow component to account for the noisy spectrum. These affect all the fitting windows and we remove them by applying an upper limit on the width of the primary components:

\begin{center}
    $\rm\sigma < 1000~km~s^{-1}$ for \sii, \nii, \oiii, $\ha$, or $\hb$
\end{center}

These quality cuts result in 1,350,112 sources across the stellar mass and redshift range, which we use for our analysis.

\subsection{Selecting Line-Emitting Galaxies} \label{subsec:candidates}
We explore two complementary ways of detecting AGN candidates in this paper: emission-line ratios and broad $\ha$ detection (Section~\ref{sec:agn_selection}). This depends on the accurate detection of four emission lines: $\hb$, \oiii$\lambda$5007, $\ha$, and \nii$\lambda$6583. We apply the following SNR and AoN (amplitude-over-noise) cuts on these four lines for selecting the sample of line-emitting galaxies:

\begin{center}
    SNR $\ge$ 3 for \oiii, $\ha$, \nii \\
    (SNR $\ge$ 1) \& (AoN $\ge$ 1) for $\hb$ \\
\end{center}

It is often found that $\hb$ lacks significant detection in galaxies and setting a higher threshold often removes good candidates that would have been classified as AGN \citep{CidFernandes+2010}. We therefore use a lower threshold for $\hb$ compared to the other emission lines. For double-peaked sources, we consider the fluxes as the sum of the fluxes of the two components, and their errors are added in quadrature for computing the SNR values. We still consider the AoN of the primary $\hb$ component for these sources as a way of ensuring that the components are detected over the noise of the spectrum. We have 459,208 sources that satisfy the SNR criteria for \oiii, $\ha$, and \nii, of which 48,451 sources are rejected based on the $\hb$ cuts. This results in 410,757 line-emitting galaxies, of which 114,496 are dwarf galaxies (51.6\%), and the rest of 296,261 (26.3\%) are high-mass galaxies. We observe broad $\ha$ ($\ha$;b) detection in 13,526 of these sources. To select confident $\ha$;b candidates, we apply the following cuts:

\begin{center}
    SNR ($\ha$;b) $\ge$ 3 \\
    SNR ($\sigma_{\ha;b}$) $\ge$ 3 \\ 
    AoN ($\ha$;b) $\ge$ 2
\end{center}

Finally, we apply a cut on {\tt PROB\_BROAD} for selecting statistically likely broad component detections:

\begin{center}
    {\tt PROB\_BROAD} $\ge$ 80\%
\end{center}

These cuts result in 6,185 broad-line (BL) candidates, and the remaining 404,572 line-emitting galaxies are considered narrow-line (NL) sources. 

\begin{figure*}
    \centering
    \includegraphics[width = \textwidth]{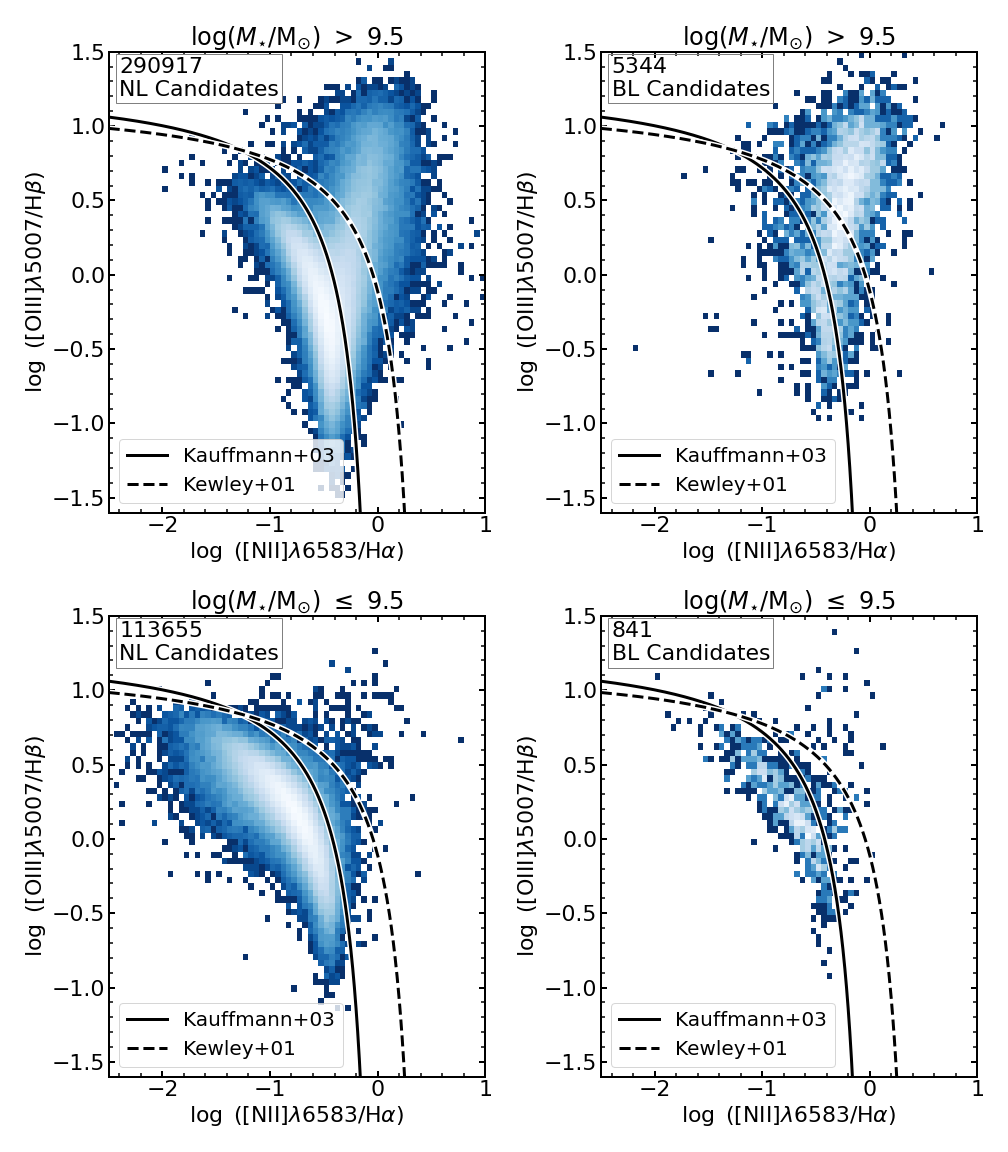}
    \caption{BPT \oiii/$\hb$ vs \nii/$\ha$ narrow-line diagnostic diagram for high-mass ($\logmass > 9.5$; {\it Top}) and dwarf ($\logmass \leq 9.5$; {\it Bottom}) galaxies, divided as NL ({\it Left}) and BL ({\it Right}) candidates. The solid line in all the panels is from \citet{Kauffmann+2003}, which separates the pure star-forming sources and those with AGN contribution. The dashed line represents the ``maximum starburst line'' using stellar photoionization models \citep{Kewley+2001}. This diagnostic diagram identifies AGN in 75,928/296,261 ($\approx$25.6\%) high-mass and 2,444/114,496 ($\approx$2.1\%) dwarf galaxies.}
    \label{fig:bpt_all}
\end{figure*}

\section{Incidence of AGN in Line-Emitting Galaxies} \label{sec:agn_selection}

In this section, we search for AGN signatures in our sample of line-emitting high-mass and dwarf galaxies (Section~\ref{subsec:bpt_diagram}). 
We further compare our identified dwarf AGN candidates with results from prior single-fiber spectroscopic surveys (Section~\ref{subsec:dwarf_agn}). Lastly, we explore the variation of the observed AGN fraction with stellar mass and redshift (Section~\ref{subsec:agn_frac}). 

\subsection{AGN Candidates from the \nii-BPT Diagram \label{subsec:bpt_diagram}} 
Two primary processes contribute to the photoionization of the interstellar medium (ISM) in galaxies: star formation and AGN activity. AGN produce a harder spectrum (i.e., a larger fraction of high-energy photons) than hot stars, causing identifiable signatures in the resulting emission line spectrum. Two-dimensional emission-line diagnostic diagrams, commonly known as the BPT diagrams \citep{bpt, bpt_vo87}, can be used to distinguish the primary source of ionization in galaxies and identify AGN. In this paper, we focus on the most widely used BPT diagram with \nii/$\ha$ vs. \oiii/$\hb$ (hereafter \nii-BPT Diagram). We explain our choice to use only this specific diagnostic in Section~\ref{subsec:sii-bpt}.

Using pure stellar photoionization models, \citet{Kewley+2001} suggested a ``maximum starburst line'' (dashed line in all panels of Figure~\ref{fig:bpt_all}), above which a non-stellar ionization source, likely an AGN, is necessary to account for the emission-line ratios. The solid line in all the panels of Figure~\ref{fig:bpt_all} separates the sources with AGN contribution from the empirical branch attributed to star-forming galaxies \citep{Kauffmann+2003}. Galaxies located between these two lines contain contributions from both AGN and star formation and are typically referred to as ``composite'' galaxies. For this study, we classify all sources in the AGN-dominated and composite regions as AGN candidates. 

The \nii-BPT diagnostic is metallicity-sensitive, with lower metallicity systems occupying regions with higher \oiii/$\hb$ and lower \nii/$\ha$ ratios \citep{Storchi-Bergmann+1998, Carvalho+2020}. Consequently, low metallicity AGN can occupy the same regions as low metallicity starbursts \citep{Groves+2006}.

\subsubsection{Initial BPT Selection \label{subsubsec:initial_bpt}}

We categorize the line-emitting galaxies in our sample into two groups: those exhibiting only narrow components (NL candidates) and those with a secure broad $\ha$ detection (BL candidates). While our primary goal is to identify AGN candidates, we also aim to separate the BL candidates where the presence of a broad $\ha$ component is indicative of the broad-line region (BLR) associated with an AGN. This will facilitate the estimation of their BH masses (Section~\ref{subsec:bh_masses}) and the study of the $\mbh - \mstar$ scaling relation (Section~\ref{subsec:mbh_mstar}). Thus, we focus on identifying AGN and composite sources in both NL and BL candidates (Section~\ref{subsec:candidates}) via the BPT diagnostic diagram.

Figure~\ref{fig:bpt_all} shows the \nii-BPT diagram of high-mass galaxies ({\it Top Panel}) and dwarf galaxies ({\it Bottom Panel}) in our sample. The left panels show the BPT diagram for the NL candidates, while the right panels show the BPT diagram for the BL candidates constructed using the line ratios of the narrow components of the emission lines. 

The shape of the bivariate distributions on the BPT diagram is different for dwarf and high-mass galaxies. The high-mass galaxies are characterized by a thin star-forming branch and a well-extended AGN branch for both NL and BL candidates. On the other hand, low-mass galaxies have a dense star-forming branch that extends to lower \nii/$\ha$ ratios ($\rm < -1.4$) than their counterparts. The dwarf galaxy diagrams also have sparsely populated AGN branches for both NL and BL candidates. This is not surprising as dwarf galaxies are typically star-forming galaxies and their star-forming signatures can therefore dominate the emission-line signatures. In addition, the low metallicity of dwarf galaxies leads to low \nii~and high \oiii~values \citep{Madden+2019, Henkel+2022}, which explains the concentration of dwarf galaxies at the top-left corner of the star-forming branch \citep{Storchi-Bergmann+1998, Carvalho+2020}. 

Of the 296,261 high-mass line-emitting galaxies (both NL and BL candidates) in our sample, we find that 24,227 (8.2\%) sources are AGN-dominated, 51,701 (17.4\%) are composites, and 220,333 (74.4\%) are star-forming galaxies. In contrast, we find 405/114,496 (0.3\%), 2,039/114,496 (1.8\%), and 112,052 (97.9\%) of the dwarf line-emitting galaxies (both NL and BL candidates) reside on the AGN-dominated, composite, and star-forming branch, respectively. Table~\ref{tab:bpt-numbers} describes the number of NL and BL candidates in each of the BPT regions from Figure~\ref{fig:bpt_all}.

\begin{deluxetable*}{lcccc}
\tablecaption{Number of AGN, Composites, and Star-Forming Candidates from the BPT Diagram}
\label{tab:bpt-numbers}
\tablehead{
\colhead{} & \multicolumn{2}{c}{\bf \underline{Dwarf ($\logmass \le 9.5$) Galaxies}} & \multicolumn{2}{c}{\bf \underline{High-Mass ($\logmass > 9.5$) Galaxies}} \\
All Sources & \multicolumn{2}{c}{221,778} & \multicolumn{2}{c}{1,128,334} \\
\hline
\colhead{} & \colhead{\bf NL Candidates} & \colhead{\bf BL Candidates} & \colhead{\bf NL Candidates} & \colhead{\bf BL Candidates}
}
\startdata
\multicolumn{5}{c}{{\bf Before Visual Inspection}\tablenotemark{a}} \\
\hline
Line Emitting Galaxies & 113,655 & 841 & 290,917 & 5,344 \\
AGN Dominated Sources & 350 (0.3\%) & 55 (6.5\%) & 21,322 (7.3\%) & 2,905 (54.4\%) \\
Composites & 1,933 (1.7\%) & 106 (12.6\%) & 50,330 (17.3\%) & 1,371 (25.6\%) \\
Star-Forming & 111,372 (98.0\%) & 680 (80.9\%) & 219,265 (75.4\%) & 1,068 (20.0\%) \\
\hline
\multicolumn{5}{c}{{\bf After Visual Inspection}\tablenotemark{b}} \\
\hline
AGN Dominated Sources & 365 (0.3\%) & 40 (4.8\%) & 21,405 (7.4\%) & 2,822 (53.6\%) \\
Composites & 1,951 (1.7\%) & 88 (10.7\%) & 50,470 (17.3\%) & 1,231 (23.6\%) \\
\enddata
\tablenotetext{a}{These numbers reflect NL and BL candidates selected based on initial criteria outlined in Section~\ref{subsec:candidates}. The AGN and composite counts are based on the narrow emission-line ratios plotted on the BPT diagram.}
\tablenotetext{b}{These numbers reflect the results after visually inspecting BL candidates with FWHM($\ha$;b) $< 1000 \, \rm km \, s^{-1}$ that fall in the AGN-dominated and composite regions of the BPT diagram (see Section~\ref{subsubsec:vi_blcand}). Sources failing the visual inspection of broad components remain classified as AGN and composites but are now categorized as NL-AGN sources.}
\end{deluxetable*}

\subsubsection{Visual Inspection of BL Candidates} \label{subsubsec:vi_blcand}

Of the total 6,185 BL candidates in our sample (both high-mass and dwarf galaxies), 1,748 (28.2\%) lie on the star-forming branch, 1,477 (23.9\%) are composites, and 2,960 (47.9\%) are AGN-dominated galaxies (see Table~\ref{tab:bpt-numbers}). The broad $\ha$ emission in the BL candidates can originate from several processes within the galaxy. For instance, infalling gas in the broad-line region (BLR) of the AGN can produce this emission. Additionally, dynamics in the accretion disk surrounding the BH, outflows driven by AGN activity, and/or star formation can contribute extra components to the $\ha$ emission line. Stellar processes, such as Type II supernovae and Luminous Blue Variables (LBVs), may also generate broad $\ha$ emission \citep{Smith+2011, Gutierrez+2017, Pessi+2023}. To distinguish between the star-formation-driven and AGN-driven broad $\ha$ sources, we focus exclusively on the BL candidates identified as AGN and composites in the \nii-BPT diagram. 

Since we push to lower widths of the broad $\ha$ component than any prior studies \citep{Reines+2013, Reines&Volonteri2015, Suh+2020, Salehirad+2022}, we aim to ensure the confident identification of such low-width broad components. For this purpose, we visually inspect 556 AGN candidates that have $\rm FWHM_{\ha;b}$ $<~\rm 1000~km s^{-1}$ (Appendix~\ref{app:vi}). 

After a thorough visual inspection, we identify 146 sources exhibiting a clear broad $\ha$ component. We consider these, along with the rest of the BL AGN-dominated and composite candidates with $\rm FWHM_{\ha;b}$ $>~\rm 1000~km s^{-1}$ as {\it confident} BL-AGN candidates. Additionally, we classify 154 sources with a statistical broad $\ha$ component, as {\it tentative} BL-AGN candidates, although their broad components are less clear visually. 

Of the remaining 256 visually inspected BL candidates, we identify 252 sources that display a visual second/outflow component in the \nii~line but are not detected by {\tt EmFit}. We suspect this explains the additional $\ha$ component rather than a true broad component. Additionally, four sources exhibit particularly complex emission line profiles around the \nii~+~$\ha$ emission lines, with multiple overlapping components, preventing us from constraining the presence of a broad component. Consequently, we do not classify these 256 galaxies as BL-AGN and relabel them as NL-AGN. Even though we do not visually inspect any BL candidates that are present on the star-forming branch, it does not affect the percentages reported in Section~\ref{subsubsec:initial_bpt} or any subsequent analysis in this paper. We present details of our visual inspection in Appendix~\ref{app:vi}.

\begin{figure*}
    \centering
    \includegraphics[width = 1.0\textwidth]{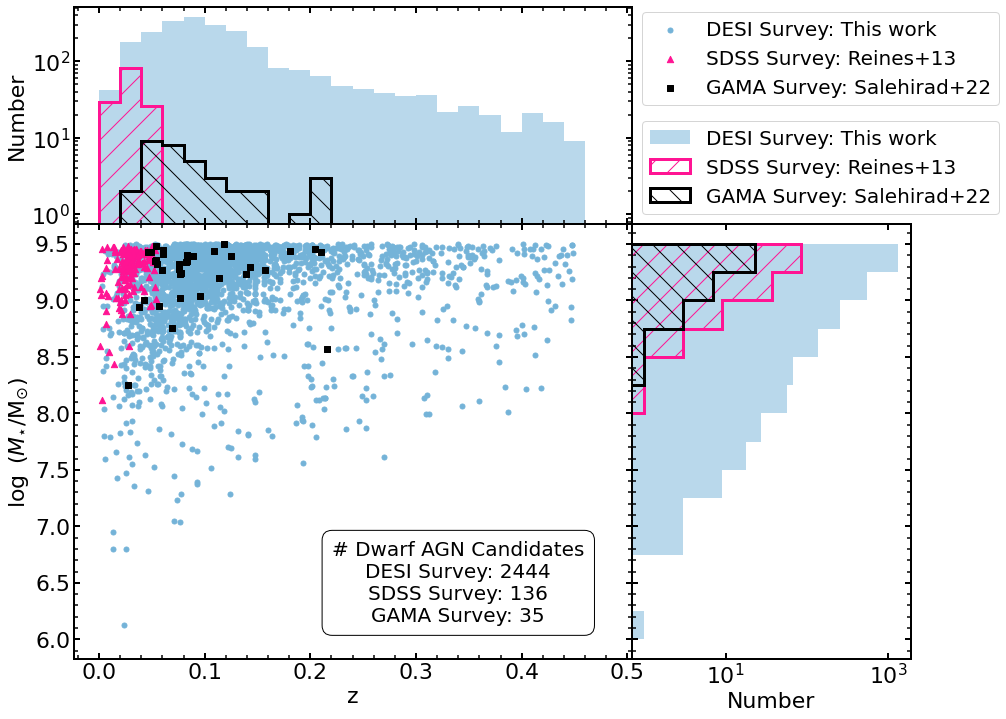}
    \caption{Comparison of dwarf AGN candidates detected from DESI with detections from previous single-fiber spectroscopic surveys. {\it Main panel:} Distribution of dwarf AGN candidates in the $\logmass - \rm z$ space. Candidates from the DESI survey (this work) are shown as blue points, while those from SDSS and GAMA surveys are shown as pink triangles and black squares, respectively. The number of dwarf AGN candidates detected in each survey is noted on the plot. The dwarf AGN candidates from this work more than triple the current census of optically identified dwarf AGN candidates. {\it Top Panel:} Redshift distribution of the dwarf AGN candidates. {\it Right Panel:} Stellar mass distribution of the dwarf AGN candidates. The histogram for DESI candidates is shown in blue, while those from SDSS and GAMA are displayed in hatched pink and hatched black, respectively. DESI significantly extends the search for dwarf AGN candidates to lower galaxy masses and higher redshifts.}
    \label{fig:mstar_z}
\end{figure*}

\subsubsection{Final BPT Selection} \label{subsubsec:final_bpt}

Of the 296,261 high-mass line-emitting galaxies, we have a final sample of 71,875 NL-AGN and 4,053 BL-AGN candidates (AGN Fraction $\approx$ 25.6\%). These numbers are influenced by the selection criterion of line-emitting galaxies (Section~\ref{subsec:candidates}). Given that robust detection of emission lines is required to appear on the BPT diagram, some high-mass quiescent galaxies hosting an AGN may be missed by this diagnostic. The spectra of such galaxies are typically dominated by strong absorption lines from stars, weakening the emission-line contribution from the AGN and resulting in a drop of high-mass AGN candidates detected via the \nii-BPT diagram. 

Of the 114,496 dwarf line-emitting galaxies, we have a final sample of 2,316 NL-AGN and 128 BL-AGN candidates (AGN Fraction $\approx$ 2.1\%). The majority of dwarf galaxies reside on the star-forming branch. This does not necessarily mean that no AGN is present in these sources. The high star formation in low-mass galaxies can dilute the contribution from AGN \citep{Trump+2015}. There might be a few dwarf AGN candidates on the star-forming branch that may be detected via other diagnostics \citep{Molina+2021, Salehirad+2022}. In fact, \citet{Birchall+2020} found that 85\% of their X-ray-confirmed dwarf AGN candidates fail the BPT criterion for AGN. However, the dwarf AGN candidates that are detected from the \nii-BPT diagram have been observed to be robust AGN candidates \citep{Reines+2013, Baldassare+2020, Salehirad+2022}. We note that we are attempting to detect confident AGN candidates in our sample of galaxies, but do not claim to be complete. Our estimates therefore provide lower limits to the incidence of AGNs in galaxies. Table~\ref{tab:bpt-numbers} provides more detailed information regarding our emission-line classification. 

\subsection{AGN in Line Emitting Dwarf Galaxies} \label{subsec:dwarf_agn}

Our result of 2,444 dwarf AGN candidates more than triples the existing census of optically selected dwarf AGN candidates to date, and is based on using just the SV data and 20\% of DESI Year 1 data. By extrapolating to the entire DESI footprint, we expect to conservatively find $\gtrsim$10,000 dwarf galaxies with detectable AGN signatures within the redshift range of 0.001 $\le\rm z\le$ 0.45, after the completion of the five-year survey.

The observed dwarf AGN fraction in this work ($\approx$2.1\%) is higher than any previous study using single-fiber integrated optical spectra \citep[AGN fraction $<$1\%;][]{Reines+2013, Salehirad+2022} or using infrared and X-ray searches \citep[AGN Fraction $<$1\%;][]{Mezcua+2018, Lupi+2020, Birchall+2020, Latimer+2021a, Birchall+2022, Bykov+2024}. In this subsection, we compare our dwarf AGN candidates with those detected using the BPT diagnostics from prior single-fiber spectroscopic surveys i.e., the Sloan Digital Sky Survey \citep[SDSS;][]{sdss} and Galaxy and Mass Assembly \citep[GAMA;][]{gama} survey. 

Figure~\ref{fig:mstar_z} shows the distribution of dwarf AGN candidates in our sample (blue circles), alongside candidates identified by \citet{Reines+2013} (with SDSS; pink triangles) and \citet{Salehirad+2022} (with GAMA; black squares) in the $\logmass~-~\rm z$ space. The top and right panels display the distribution of redshift and stellar masses of these samples, respectively. All stellar masses have been estimated using the Chabrier IMF. 

\citet{Reines+2013} conducted the first systematic search for AGN in dwarf galaxies using the SDSS DR8 spectroscopic catalog \citep{sdss_dr8}, covering an area of $\approx\rm9200~deg^{-2}$. Using the \nii-BPT emission-line diagnostic, they found 136 candidates with AGN signatures down to $\logmass = 8.1$ and out to z = 0.055, with a median $\logmass = 9.3$ and a median redshift of z = 0.028. More recently, \citet{Salehirad+2022} applied various spectroscopic diagnostics on galaxies from the GAMA DR4 \citep{gama_dr4}, spread over an area of $\approx\rm250~deg^{-2}$. They found 35 dwarf AGN candidates down to $\logmass = 7.99$ and out to z = 0.2. These candidates have a slightly higher median redshift (z = 0.077) compared to \citet{Reines+2013}, but have a similar median $\logmass = 9.3$.

DESI has greatly expanded the discovery space of optical dwarf AGN candidates. Our candidates extend down to $\logmass \approx$ 6.12, almost 1.5 dex lower than both SDSS and GAMA surveys, and also extend out to z $\approx$ 0.45.  The early DESI data covers a sky area of $\approx\rm4,400~deg^{-2}$, nearly half of the SDSS DR8 coverage ($\approx\rm9,200~deg^{-2}$). The larger number of dwarf AGN candidates from our study is partly due to a higher target density of DESI compared to the previous spectroscopic sky surveys. We explore the selection effects compared to \citet{Reines+2013} in more detail in Section~\ref{subsec:reines13}.

\subsection{AGN Fraction in Galaxies}\label{subsec:agn_frac}

The fraction of dwarf galaxies hosting a BH, i.e., their BH occupation fraction, is a key diagnostic in understanding the BH seed formation mechanisms \citep{Volonteri2010, Mezcua2017}. Some of these BHs can be detected as an AGN via different multi-wavelength methods. Therefore, the AGN fraction acts as a lower limit for the BH occupation fraction. We estimate the AGN fraction in DESI sources based on the fraction of galaxies identified as BPT-AGN candidates. As is common in previous optical studies of dwarf galaxies \citep{Reines+2013, Salehirad+2022, Mezcua+2020, Mezcua+2024a}, we do not apply any completeness corrections. However, we will consider the effect of accounting for the galaxies that fail the emission line detection criteria from Section~\ref{subsec:candidates}.

\begin{figure}
    \centering
    \includegraphics[width = 1.0\columnwidth]{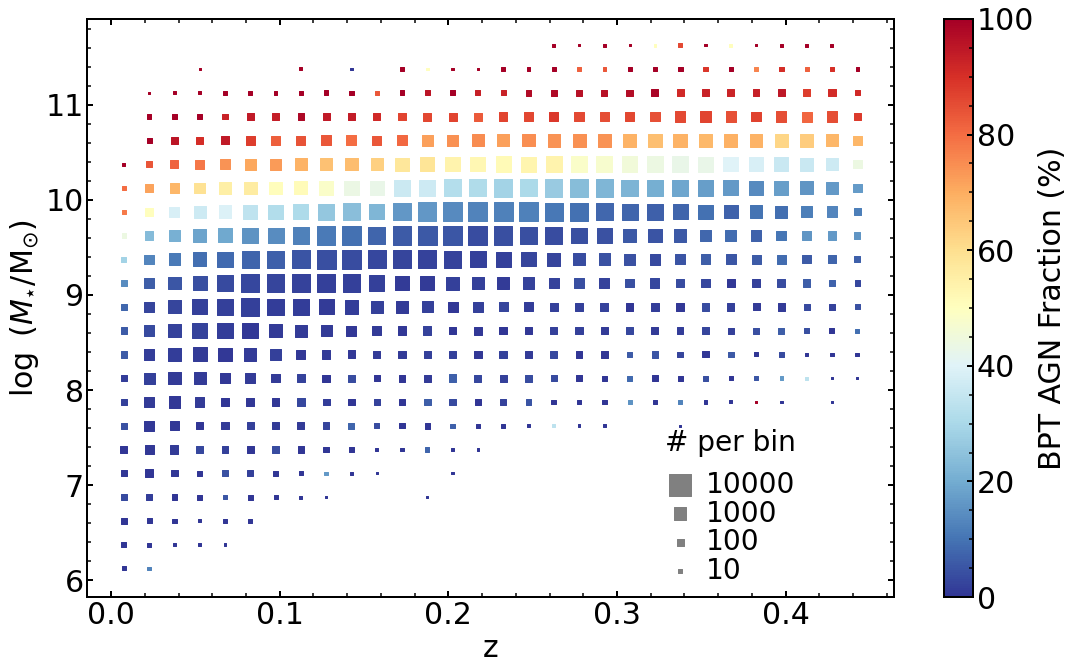}
    \caption{2D-Distribution of line-emitting galaxies in the  $\logmass - \rm~z$ space color-coded by the BPT-AGN fraction within the bins. The size of the squares is proportional to the total number of line-emitting galaxies within the bin.}
    \label{fig:mstar-z-agnfrac}
\end{figure}

To investigate trends with galaxy properties, we start by computing the AGN fraction in line-emitting galaxies in bins of stellar mass and redshift (Figure~\ref{fig:mstar-z-agnfrac}). The size of the squares shown in the figure is proportional to the number of galaxies within those bins, while their color reflects the AGN fraction as shown on the color bar. We find that the BPT-AGN fraction, on average, increases with increasing stellar mass, across the entire redshift range, which is consistent with previous optical emission-line studies \citep[e.g.,][]{Juneau+2011}. The BPT-AGN fraction is $\lesssim$10\% at low stellar masses ($\logmass \le$ 9), and reaches $\approx$100\% at high stellar masses ($\logmass \ge 11$). At $\logmass \approx 10$, the BPT-AGN fraction appears to decrease with increasing redshift. The change with redshift is unclear at lower and higher masses than this region. We discuss the variation of the BPT-AGN fraction with stellar mass and redshift, along with various selection effects below.

\begin{figure*}
    \centering
    \includegraphics[width = 1.0\textwidth]{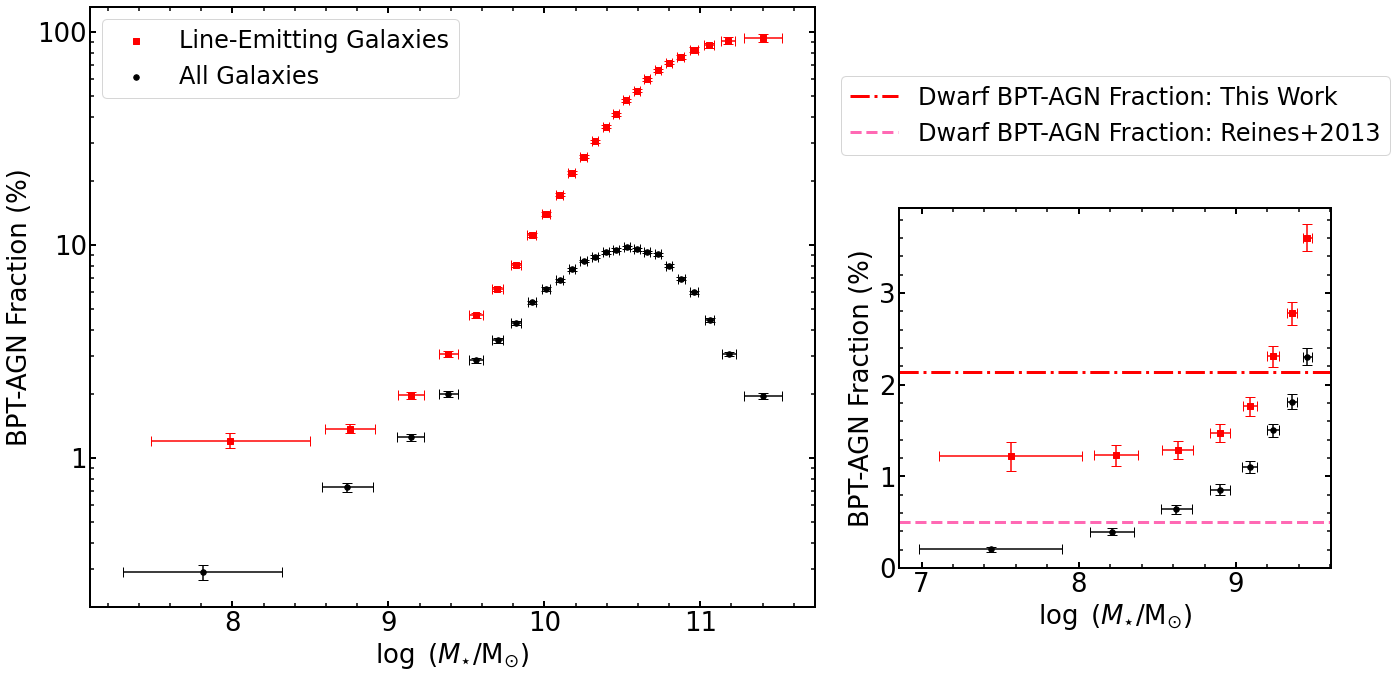}
    \caption{BPT-AGN Fraction as a function of stellar mass: The fraction of BPT-AGN candidates in line-emitting galaxies is shown as red squares, while the fraction of BPT-AGN candidates considering all galaxies is shown as black circles. {\it Right:} BPT-AGN Fraction as a function of stellar mass in the dwarf galaxy regime. The overall observed BPT-AGN fraction from our study is shown as a dashed-dotted red line, while that from \citet{Reines+2013} is shown as a dashed pink line. The BPT-AGN fraction in line-emitting galaxies increases with stellar mass, while the BPT-AGN fraction when considering all galaxies peak at $\logmass \approx 10.5$ and then declines towards higher masses. The divergence between these two estimates is primarily driven by selection biases related to the detection of emission lines.}
    \label{fig:agnfrac-mstar}
\end{figure*}

\subsubsection{Variation with Stellar Mass}

In Figure~\ref{fig:agnfrac-mstar}, we show the observed BPT-AGN fraction as a function of stellar mass. We arrange all the sources in order of increasing stellar masses and divide them into bins of nearly equal number of sources ($\approx$54,000 sources in each bin). We compute two different BPT-AGN fraction estimates: the fraction of BPT-AGN candidates in line-emitting galaxies (shown as red squares in Figure~\ref{fig:agnfrac-mstar}); and the fraction of BPT-AGN candidates in all galaxies, including the galaxies without line emission (shown as black circles in Figure~\ref{fig:agnfrac-mstar}). Even though some non-line-emitting galaxies may host a BH, we assume that they do not host an AGN. Therefore, we are reporting the lower limits of the total BPT-AGN fraction. Within each bin, we compute the median and standard deviation of stellar masses as the representative stellar mass and error. The right panel of Figure~\ref{fig:agnfrac-mstar} shows the variation of the BPT-AGN fraction in the dwarf galaxy regime. In this case, we divide all the dwarf galaxies into bins of $\approx$27,700 sources and compute both versions of the BPT-AGN fraction. The horizontal dashed-dotted red line denotes the observed average dwarf AGN fraction in line-emitting galaxies (Section~\ref{subsec:dwarf_agn}). This fraction is nearly 4 times higher than the pink dashed line that denotes the average dwarf AGN fraction from \citet{Reines+2013}.

Focusing on line-emitting galaxies, we observe the BPT-AGN fraction sharply rises from $\approx$1.2\% at $\logmass \approx 8.0$ to $\approx$93.5\% at $\logmass \approx$ 11.4. While it is well established that nearly all massive galaxies host BHs \citep{Kormendy&Ho2013}, only a subset of these BHs are actively accreting at any given time. Furthermore, strong selection effects are present in this regime, as many sources fail to be detected in all four BPT lines. As a result, there is an increasing bias towards the detection of emission lines linked to the presence of AGN with increasing stellar mass.

At the low-mass end, whether BHs are present at the centers of all dwarf galaxies has not yet been established. The BPT-AGN fraction in dwarf galaxies (right panel of Figure~\ref{fig:agnfrac-mstar}) increases from $\approx$1.2\% at $\logmass \approx 7.6$ to $\approx$3.6\% at $\logmass \approx 9.4$. It drops to $<$2.1\% (red horizontal line) by $\rm\log (\mstar/\msun) \approx 9$, which is the average dwarf AGN fraction. This suggests that the majority of dwarf AGN candidates from the BPT selection criterion are dwarf galaxies with $\logmass > 9$, which is what we see in Figure~\ref{fig:mstar-z-agnfrac}. An important selection bias in this mass regime is that for lower-mass systems, the detected AGN sources are likely accreting at higher Eddington ratios, and hence only the more extreme sources would be selected in dwarf galaxies compared to massive galaxies. 

\begin{figure}
    \centering
    \includegraphics[width = 1.0\columnwidth]{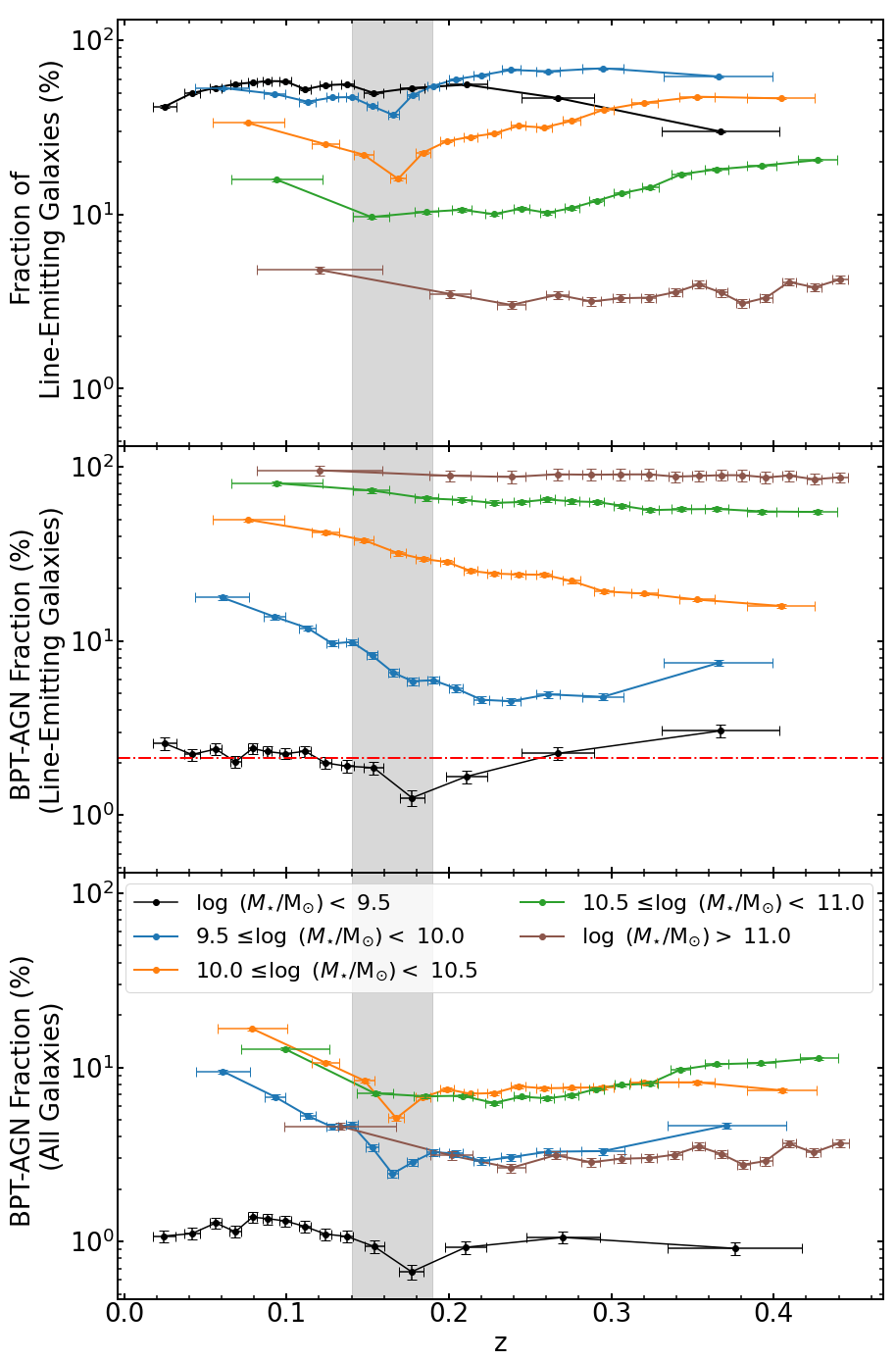}
    \caption{{\it Top Panel:} Fraction of line-emitting galaxies as a function of redshift; BPT-AGN Fraction as a function of redshift considering only line-emitting galaxies ({\it Middle Panel}) and considering all galaxies ({\it Bottom Panel}). In all panels, the fractions in dwarf galaxies are plotted in black, while the massive galaxies are divided into bins of stellar mass. The variation of BPT-AGN fraction with redshift is subject to several selection effects. The red horizontal line in the middle panel denotes the overall observed dwarf BPT-AGN fraction. The shaded gray region denotes the redshift range when the \oiii~emission line coincides with the overlap region of the DESI B and R cameras.}
    \label{fig:agnfrac-z}
\end{figure}

When we include non-line-emitting galaxies that do not have significant emission-line detection (black points in Figure~\ref{fig:agnfrac-mstar}), we observe that the BPT-AGN fraction increases from $\approx$0.3\% at $\logmass \approx 7.8$, peaks at $\approx$9.7\% at $\logmass \approx 10.5$, turnovers and decreases to $\approx$2\% at $\logmass \approx 11.4$. The discrepancy between the two estimates of the BPT-AGN fraction is due to the fraction of line-emitting galaxies selected at different stellar masses (see top panel of Figure~\ref{fig:agnfrac-z}). In the stellar mass range of $9 \le \logmass \le 10$, the difference is small, and $>$60\% of the sources have strong emission-line detection. However, there is an increase in the divergence between the line-emitting and total BPT-AGN fraction (red and black curves) at higher stellar masses. This divergence occurs because massive galaxies tend to be metal-rich and have weak \oiii~emission \citep{Kauffmann+2003}, often failing the selection criterion. In contrast, the differences in the low-mass regime are primarily driven by the \nii~selection criterion, which is expected due to the low metallicity of galaxies in this regime. 

The increase in the AGN fraction with increasing stellar mass is similar to some previous studies focusing on X-ray-detected AGN \citep{Mezcua+2018, Birchall+2020}. However, other X-ray and multiwavelength studies have found weak to no variation with stellar mass \citep{Juneau+2013, Birchall+2022, Birchall+2023}, suggesting that different AGN selection criteria have different selection biases. 

\subsubsection{Variation with Redshift}

We found that the BPT-AGN fraction strongly depends on stellar mass. To examine the variation of BPT-AGN fraction with redshift, we group all the dwarf galaxies and sub-divide high-mass galaxies into stellar mass bins of 0.5 dex. We then organize these sources in increasing order of redshift and calculate the fraction of line-emitting galaxies, as well as the line-emitting and total BPT-AGN fractions, within redshift bins containing nearly equal number of sources. The top panel of Figure~\ref{fig:agnfrac-z} illustrates the fraction of line-emitting galaxies across different stellar mass selections as a function of redshift, while the middle and bottom panels display the variations of line-emitting and total BPT-AGN fractions with redshift for these groups, respectively. The horizontal red line in the middle panel is the observed AGN fraction in dwarf galaxies. 

A noticeable dip in the estimated fractions is evident in all the panels around $\rm z\approx$ 0.18. This dip arises from two issues concerning the \oiii~emission line. In this redshift range, the \oiii~line coincides with the overlap region of the B and R cameras of the DESI spectrograph (shown by grey region in Figure~\ref{fig:agnfrac-z}), and also one of the skylines (the Na I doublet). As a result, the detection of this emission line becomes very noisy, often resulting in the SNR (\oiii) failing to meet the selection criterion in this redshift range.  

In the dwarf galaxy regime (black line), the fraction of line-emitting galaxies remains relatively constant at $\approx$50\% up to $\rm z \approx$ 0.2, before declining to $\approx$20\% at $\rm z \approx$ 0.37. The BPT-AGN fraction for line-emitting dwarf galaxies fluctuates around the overall dwarf AGN fraction ($\approx$2.1\%; dash-dotted line) and remains nearly constant throughout the entire redshift range. When considering all dwarf galaxies and assuming that the non-line-emitting galaxies do not host an AGN, the BPT-AGN fraction drops to $\lesssim$ 1\% but remains nearly flat with redshift. Thus, the dwarf AGN fraction is almost a factor of two higher when considering only line-emitting galaxies versus all galaxies but with no significant redshift dependency. 

When we focus on massive galaxies (colored lines), we find a clear dependence of the trends on stellar mass. As stellar mass increases, the fraction of line-emitting galaxies decreases across all redshifts. Conversely, the BPT-AGN fraction for line-emitting galaxies (middle panel) exhibits an opposite trend and increases with stellar mass across all redshifts. When considering all galaxies (bottom panel), there is no longer a monotonic trend of AGN fraction with stellar mass. However, the trend with redshift behaves similarly between the mass bins: we find that the total BPT-AGN fraction initially decreases, but is nearly constant at $\rm z \gtrsim$ 0.2.

Overall, the observed trends are influenced by various selection effects including the emission-line detection at different stellar masses. Moreover, the DESI targeting selection also introduces significant biases (Section~\ref{subsec:spectroscopy}). The {\tt BGS} sources are uniformly distributed across all stellar masses and redshifts. However, the rest of the high-mass galaxies are predominantly {\tt LRG} targets, which primarily target high-redshift sources. The {\tt SCND} targets also include a majority of high-mass and dwarf galaxies in our sample, but their distribution across redshifts is irregular due to different selection criteria (see Figure~\ref{fig:mstar_dist}). These variations in targeting complicate the line-emitting galaxy fraction and BPT-AGN fraction as a function of redshift.  Detailed analyses involving selection functions and incompleteness corrections are necessary for a thorough physical interpretation, which we will address in a future publication.

\section{Black Hole Mass - Stellar Mass Scaling Relation} \label{sec:scaling_relation}

BH-galaxy scaling relations hold clues about BH seed formation mechanisms and also provide a way to study the co-evolution of BHs and their host galaxies. In this section, we study the $\mbh - \mstar$ scaling relation of the BL-AGN candidates and compare it with previous studies.

\subsection{Black Hole Masses} \label{subsec:bh_masses}

The broad $\ha$ emission line can be used to compute BH masses using single-epoch virial techniques. Under the assumption that the BLR gas around the BH is virialized, the kinematics of the gas can be used as a dynamical tracer of BH mass ($\mbh$):

\begin{equation*}
    \mbh \propto \frac{RV^{2}}{G}
\end{equation*}

\noindent where $R$ is the radius of the BLR around the BH, $V$ is the average velocity of the gas, and $G$ is the gravitational constant. The average velocity of the gas can be inferred using the width of the broad $\ha$ emission line, while the radius of the BLR is approximated using the radius-luminosity relation. 

Using the same approach as previous studies \citep{Reines+2013, Reines&Volonteri2015, Salehirad+2022}, we use the \citet{Greene&Ho2005} formula for estimating the BH masses, with the modified radius-luminosity relationship of \citet{Bentz+2013} as derived by \citet{Reines+2013}:

\begin{equation}
    \begin{split}
        \log \left(\frac{\mbh}{\msun}\right) =  \log \epsilon + 6.57 + 0.47\log \left(\frac{L_{\ha;b}}{10^{42}~\rm erg~s^{-1}}\right) \\
        + 2.06\log \left(\rm \frac{FWHM_{\ha;b}}{10^{3}~km~s^{-1}}\right)
    \end{split}
\end{equation}

\noindent where $L_{\ha;b}$ and $\rm FWHM_{\ha;b}$ are the luminosity and FWHM of the broad $\ha$ component and $\epsilon$ is the scale factor, which spans a range of $\sim$0.75$-$1.4 \citep{Greene&Ho2007, Grier+2013, Grier+2017}. Here, we assume $\epsilon$ = 1 to compare with previous results \citep{Reines&Volonteri2015}.

\begin{figure}
    \centering
    \includegraphics[width = 1.0\columnwidth]{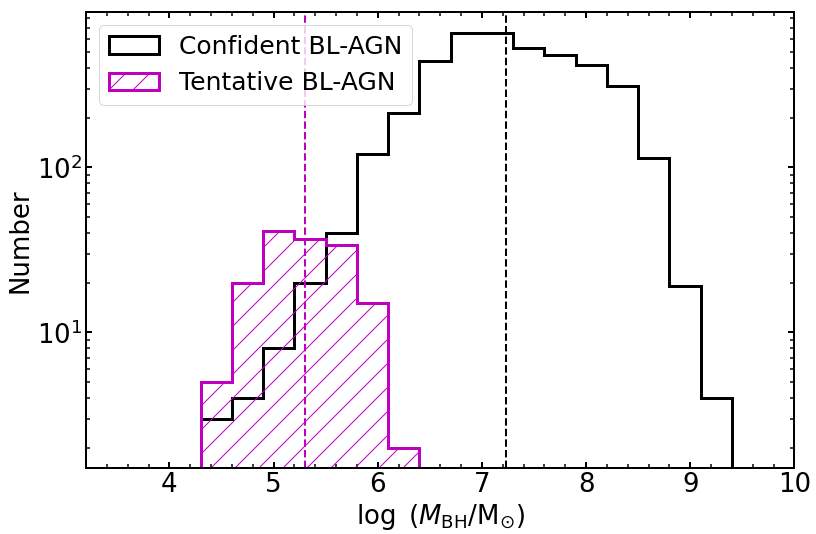}
    \caption{Distribution of BH masses of BL-AGN candidates. The {\it confident} BL candidates are shown as black histogram, while the {\it tentative} BL candidates are shown as hatched magenta histogram. The vertical black and magenta lines denote the median BH mass for the confident and tentative candidates, respectively.}
    \label{fig:bhmass_dist}
\end{figure}

We compute BH masses based on the {\tt EmFit} measurements of the broad $\ha$ flux and line width (Section~\ref{subsubsec:measurements}) and propagate their uncertainties. The median error for our sample of BL-AGN candidates is $\approx$0.02 dex. However, virial BH mass measurements from single-epoch spectroscopy can have several systematic uncertainties on the order of 0.5 dex \citep{McGill+2008, Shen2013}. We therefore add 0.5 dex to our estimated uncertainties in quadrature to get the final uncertainties in $\mbh$.

\begin{figure*}
    \centering
    \includegraphics[width = 1.0\textwidth]{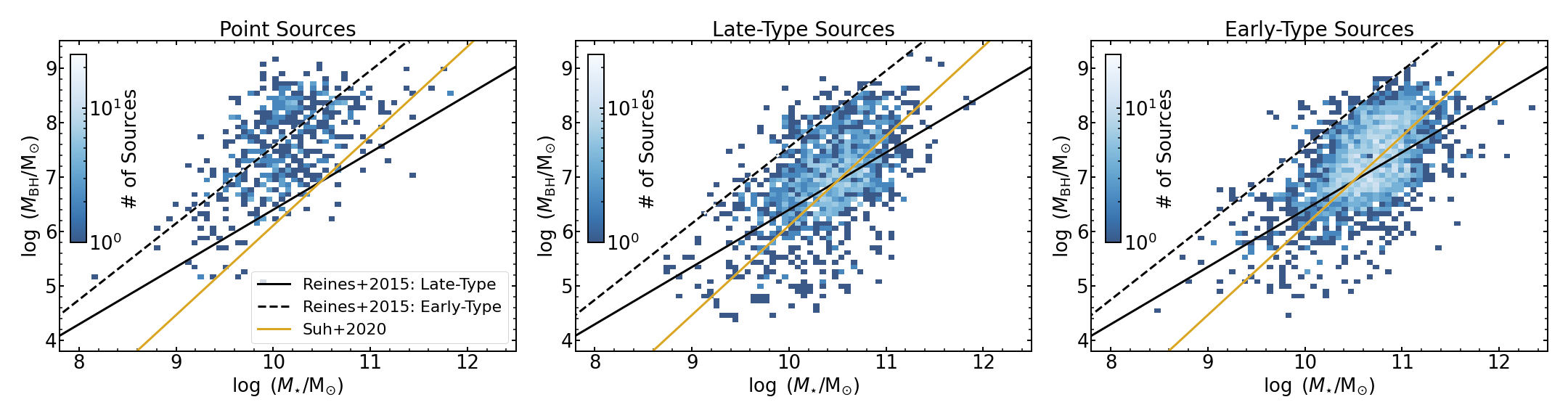}
    \caption{$\mbh - \mstar$ scaling relation of BL-AGN candidates with PSF ({\it Left}), late-type({\it Middle}), and early-type ({\it Right}) host-galaxy morphologies as defined using LS DR9. The color bars denote the number of sources within each bin of the 2D distribution. The empirical fits for late-type and early-type galaxies estimated by \citet{Reines&Volonteri2015} are shown as solid and dashed black lines, respectively. The fit from 
    \citet{Suh+2020} is plotted as an orange line.}
    \label{fig:mbh-mstar-morph}
\end{figure*}

Figure~\ref{fig:bhmass_dist} presents the distribution of BH masses for the 4,027 confident BL-AGN candidates (in black) and 154 tentative BL-AGN candidates (in hatched magenta) from our sample. The BH masses of the confident BL-AGN sources range from $\logmbh$ = 4.4 to 9.2, with a median of 7.2. On the other hand, the tentative BL-AGN candidates are concentrated at the lower end of the BH mass range, with values between $\logmbh = 4.3$ and 6.3, with a median of 5.3. 

Among the BH candidates, 151 confident and 147 tentative BL-AGN candidates have BH masses of $\logmbh \le$ 6, identifying them as potential candidates for IMBHs\footnote{Of these, 70 IMBH candidates (46 confident and 24 tentative) reside in dwarf galaxies.}. This constitutes the largest sample of IMBHs identified to date, almost doubling their existing census \citep{Chilingarian+2018, Goradzhanov+2024}. 

\subsection{Dependency on Galaxy Morphology} \label{subsec:mbh_mstar_morph} 

Combining BH masses (Section~\ref{subsec:bh_masses}) and host galaxy stellar masses (Section~\ref{subsec:mstar}), we can now build the $\mbh - \mstar$ scaling relation. First, we assess whether this relation depends on host galaxy morphology by dividing our sample of BL-AGN based on the {\tt TYPE} and {\tt SERSIC} columns from the LS DR9 catalog (Section~\ref{subsec:photometry}). Point sources are defined based on {\tt MORPHTYPE = PSF}, while the late-type sources are selected if the {\tt MORPHTYPE} is either a round-exponential ({\tt REX}) or an exponential ({\tt EXP}) or a S\'ersic ({\tt SER}) with a S\'ersic index $\le$ 2. The rest of the candidates that do not have PSF or late-type morphologies are considered early-type galaxies. We show the bivariate distribution of these sources in the $\mbh - \mstar$ space in Figure~\ref{fig:mbh-mstar-morph}, along with the empirical fits by \citet{Reines&Volonteri2015} and \citet{Suh+2020}. 

\citet{Reines&Volonteri2015} studied the $\mbh - \mstar$ scaling relation by considering 262 BL-AGNs from SDSS DR8 spectroscopic catalog, 15 reverberation-mapped AGNs from \citet{Bentz&Katz2015}, and 79 inactive galaxies with dynamical BH masses from \citet{Kormendy&Ho2013} at z $\le$ 0.055. They identified two distinct fits with different slopes for active and inactive galaxies, which correlate with galaxy morphology. Specifically, the late-type galaxies are located around the fit for active galaxies, while the early-type galaxies align with the fit for inactive galaxies. Building on this, \citet{Suh+2020} combined the local AGN sample from \citet{Reines&Volonteri2015} with an additional 100 BL-AGNs from the COSMOS field, extending their analysis to z $\approx$ 2.5. They find a comparatively steeper fit, with their high-redshift sample predominantly consisting of high stellar mass ($\logmass \gtrsim$ 11) and high BH mass ($\logmbh \gtrsim$ 7) sources.

From Figure~\ref{fig:mbh-mstar-morph}, we find that the late-type and early-type galaxies occupy similar regions in the $\mbh - \mstar$ space, while the point sources lie $\approx$1 dex above them. Both the distributions for $\logmass \ge$ 10 are consistent with active galaxy fit (late-type fit) from \citet{Reines&Volonteri2015}, with the sources with $\logmbh \le$ 6 falling below it. 

\begin{figure*}
    \centering
    \includegraphics[width = 1.0\textwidth]{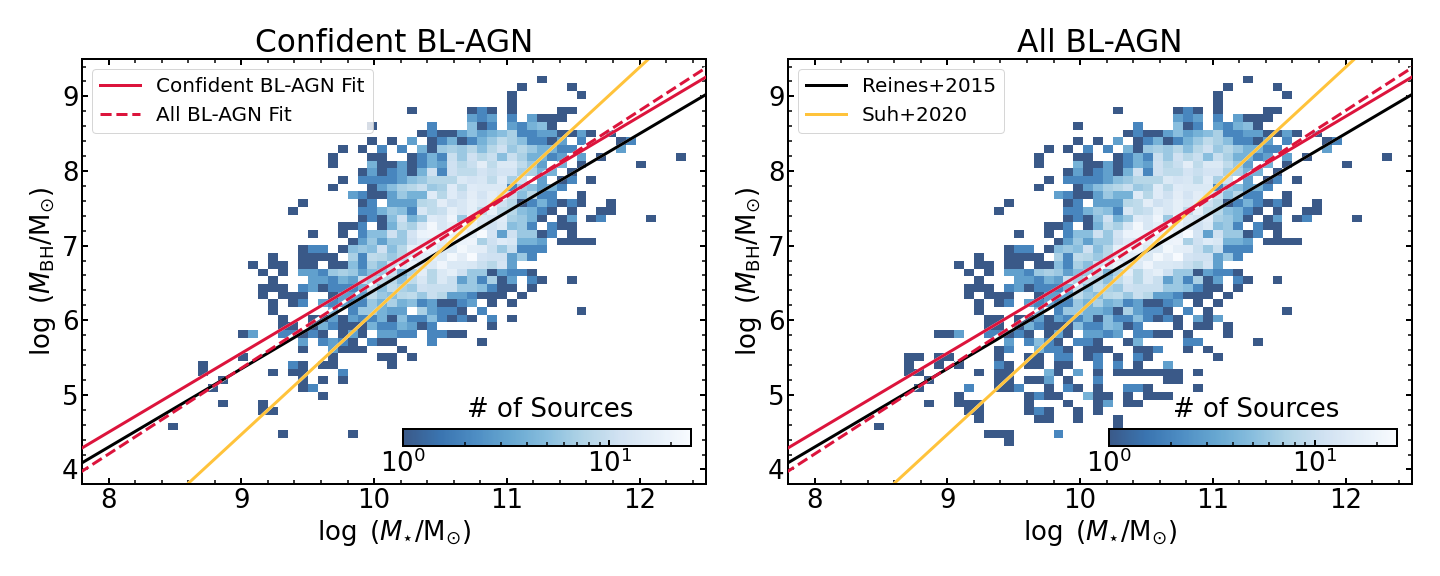}
    \caption{$\mbh - \mstar$ scaling relation of {\it confident} BL-AGN candidates ({\it Left}) and all BL-AGN candidates({\it Right}), excluding point sources. The color bar represents the number of sources within each bin of the 2D distribution. In both panels, our empirical fits for confident and all BL-AGN candidates are shown as solid and dashed red lines, respectively. The fits from \citet{Reines&Volonteri2015} and \citet{Suh+2020} are shown as black and orange lines, respectively. The empirical fits are in good agreement with the local relationship from \citet{Reines&Volonteri2015}, with the majority of the tentative BL-AGN candidates lying below the fit.}
    \label{fig:mbh-mstar}
\end{figure*}

Galaxies with a point source morphology lie toward higher values of BH masses and do not appear to follow the late-type (active) fit from \citet{Reines&Volonteri2015} nor the \citet{Suh+2020} fit. However, the early-type galaxy fit from \citet{Reines&Volonteri2015} overlaps with the distribution of these sources. Visual inspection of their spectra reveals quasar-like characteristics, featuring extremely broad emission lines and weak to negligible stellar continuum. This makes their stellar mass estimates highly uncertain. Keeping this limitation in mind, we find that 43 dwarf galaxies with PSF morphologies host BHs with $\logmbh \gtrsim$ 5. If their stellar masses were accurate or possibly over-estimated, these sources could represent low-redshift counterparts to the over-massive BHs identified by JWST \citep{Harikane+2023, Maiolino+2023, Ubler+2023}. We discuss this further in Section~\ref{subsec:bh_seeds}. 

To analyze and find empirical fits to the $\mbh - \mstar$ scaling relation, we exclude the 518 point sources due to the uncertainty with their stellar mass estimates. Given that the rest of the BL-AGN candidates are consistent with the \citet{Reines&Volonteri2015} fit based on active galaxies, we will focus solely on comparing our empirical fit to this late-type (active) fit in the following subsection. 

\subsection{Empirical Fit to the $\mbh - \mstar$ Scaling Relation} \label{subsec:mbh_mstar} 

We use the 3,633 BL-AGN candidates (3,517 confident and 146 tentative), excluding the point sources, to study the $\mbh - \mstar$ scaling relation. Figure~\ref{fig:mbh-mstar} shows the bivariate distribution of the confident BL-AGN (left panel) and all BL-AGN sources (including tentative candidates; right panel) in the $\mbh - \mstar$ space.

Focusing on confident BL-AGN candidates, we observe a significant spread in BH masses for a given stellar mass (nearly 3 dex at $\logmass \approx$10.5). Compared to previous local measurements of the relation \citet{Reines&Volonteri2015, Greene+2020}, the scaling relation now extends to lower galaxy and BH masses, down to $\logmass \approx$ 8.5 and $\logmbh \approx$ 4.4. When we include the tentative candidates, we also see a similar spread in BH masses at lower mass galaxies ($\approx$3 dex at $\logmass \approx$9.2). We find a total of 86 dwarf BL-AGN  (64 confident and 22 tentative) candidates, extending up to $\rm z \le$ 0.45. These together represent 5 times more candidates than the 14 low-mass candidates presented in \citet{Reines&Volonteri2015}. 

To study the correlation between $\mbh$ and $\mstar$, and to compare with previous studies, we fit a linear model for the two samples separately, using the Bayesian approach by \citet{Kelly2007}{\footnote{\url{https://github.com/jmeyers314/linmix}}}. We parametrize the relation similar to \citet{Reines&Volonteri2015}, as follows:

\begin{equation}
    \log (\mbh/\msun) = \alpha + \beta \log (\mstar/10^{11} \msun)
\end{equation}

Considering only the confident BL-AGN candidates, we find:

\begin{equation}
    \alpha = 7.67 \pm 0.01; \beta = 1.05 \pm 0.02
\end{equation}
    
\noindent When we consider all the BL-AGN candidates, we find:

\begin{equation}
    \alpha = 7.66 \pm 0.01; \beta = 1.15 \pm 0.03
\end{equation}

We overplot these fits as solid and dashed red lines, respectively, in Figure~\ref{fig:mbh-mstar}. We see that including the tentative candidates leads to a slightly steeper slope but with a similar normalization compared to the fit with only confident candidates. We find that our fits exhibit a similar slope but a higher normalization compared to the \citet{Reines&Volonteri2015} fit (black line). In contrast, the fit from \citet{Suh+2020} (orange line) has a steeper slope relative to both our fit and the \citet{Reines&Volonteri2015} fit. The primary divergence arises at the high-mass end, which is influenced by selection biases. Specifically, \citet{Suh+2020} focuses on massive BHs selected based on their strong X-ray emission and extremely broad $\ha$ lines, while \citet{Reines&Volonteri2015} do not include such sources in their sample. The closest counterparts of such sources in our sample are the point sources that have high BH masses (left panel of Figure~\ref{fig:mbh-mstar-morph}) and have been excluded from our bestfit.\footnote{If we assume the stellar masses of these point sources are accurate, we find a fit consistent with that of the confident candidates ($\alpha = 7.66 \pm 0.01; \beta = 1.02 \pm 0.03$). However, if some of the stellar masses of these point sources are underestimated, by more than 1 dex, this could potentially lead to a steeper fit that aligns more closely with the \citet{Suh+2020} fit.}

Excluding the point sources removes candidates with potentially over-massive BHs that are being ubiquitously discovered at high redshifts by JWST \citep{Harikane+2023, Maiolino+2023, Greene+2024}. However, we find potential under-massive BHs near $9 \le \logmass \le 11$ and $4.4 \le \logmbh \le 6$ (right panel of Figure~\ref{fig:mbh-mstar}). This scenario is possible if the BH growth is inefficient or delayed in these galaxies. The presence of star formation in low-mass galaxies can stunt the growth of BHs due to supernova feedback heating the gas and inhibiting accretion. This feedback can further hinder the gas from getting to the nuclear regions as a result of the low gravitational potential of these sources \citep{Dubois+2015, Bryne+2023}. A possibility for such under-massive BHs in massive galaxies may be associated with recent mergers, where an SMBH is merging with a lower-mass BH \citep{Kelley2021}. In such instances, the lower-mass BH could be undergoing accretion \citep{DOrazio+2023}, causing us to observe the broad $\ha$ component from its BLR kinematics. The main uncertainty in these sources comes from the intrinsic difficulty in differentiating whether the line widths are due to AGN or outflows (Appendix~\ref{app:vi}). However, excluding or including them does not significantly affect our fit.

While the extension of the $\mbh - \mstar$ scaling relation to lower galaxy masses may provide clues regarding the origin of BH seeds \citep{Mezcua2017, Volonteri2010}, it can also point to the variations in the modes of accretion in galaxies \citep{Ricarte+2018}. We explore this connection of the scaling relation with theoretical models in Section~\ref{subsec:bh_seeds}. 

\section{Discussion} \label{sec:discussion}
In this section, we examine how our main results on the AGN fraction in dwarf galaxies, BH masses, and the $\mbh - \mstar$ scaling relation may be impacted by our choices and assumptions. We describe our reasoning for using only the \nii-BPT diagram in Section~\ref{subsec:sii-bpt}. We investigate the higher dwarf AGN fraction observed from DESI compared to SDSS in Section~\ref{subsec:reines13}. Finally, we compare our $\mbh - \mstar$ scaling relation and its relation to the BH seed formation models in Section~\ref{subsec:bh_seeds}.

\subsection{\sii-BPT Diagram} \label{subsec:sii-bpt}

The \nii-BPT diagram is a valuable diagnostic for selecting AGN-dominated sources, but it is metallicity sensitive \citep{Storchi-Bergmann+1998, Carvalho+2020}. This may be problematic for dwarf galaxies, which generally have low metallicities. As a result, low-metallicity star-forming and AGN dwarfs can occupy the same region as low-mass starbursts. 

Apart from the \nii-BPT diagram, two other emission-line diagnostics are often used in identifying AGN signatures in galaxies: \sii/$\ha$ vs. \oiii/$\hb$ (\sii-BPT Diagram) and \oi/$\ha$ vs. \oiii/$\hb$ (\oi-BPT Diagram). These two diagnostics reduce the effect of metallicity and separate the star-forming galaxies from AGN-dominated Seyferts and low-ionization nuclear-emitting regions (LINERs). \citet{Polimera+2022} revised the AGN selection criterion in dwarf galaxies using all three BPT diagrams. They included a new class of dwarf galaxies, the star-forming AGN (SF-AGN) that lie on the star-forming branch in the \nii-BPT diagram, but are Seyferts/LINERs from \sii~or \oi-BPT diagrams. Using this new classification scheme for dwarf galaxies, they find an AGN fraction, ranging from $\approx$3\% to $\approx$16\% depending on the catalog used. More recently, \citet{Mezcua+2024a} also found a higher AGN fraction $\approx$20\% when combining all three diagnostic diagrams. However, the validity of the current demarcation lines is unclear and there have been several efforts to improve them \citep{Ji+2020, Law+2021}. In this subsection, we examine the demarcation lines used for the \sii-BPT diagrams. 

\begin{figure}
    \centering
    \includegraphics[width = 1.0\columnwidth]{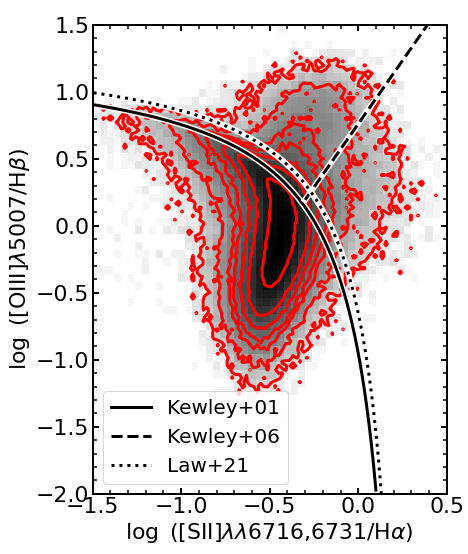}
    \caption{\sii-BPT Diagram of all the sources in our sample that pass the required criteria (see text). The contours of the source density distribution are shown in red. The solid black line is from \citet{Kewley+2001} that separates star-forming galaxies from Seyferts/LINERs. The dashed black line is from \citet{Kewley+2006} that separates Seyferts from LINERs. The new demarcation line that separates star-forming sources from Seyferts/LINERs by \citet{Law+2021} is shown as a dotted black line. The observed AGN fraction is sensitive to the chosen demarcation line.}
    \label{fig:sii-bpt}
\end{figure}

We start with the 1,350,112 sources in our sample with robust stellar masses and emission-line measurements in the redshift range of interest (Section~\ref{sec:methods}). We apply the following SNR and AoN cuts on the four emission lines required for the \sii-BPT diagram:

\begin{center}
    SNR $\ge$ 3 for \sii, $\ha$, \oiii \\
    (SNR $\ge$ 1) \& (AoN $\ge$ 1) for $\hb$ \\
\end{center}

These cuts result in 449,295 sources with $\logmass \ge 6.0$. Figure~\ref{fig:sii-bpt} shows the \sii-BPT diagram of all these sources. The solid black lines in the figure denote the demarcation line for separating star-forming galaxies from the AGN-dominated Seyferts and LINERs \citep{Kewley+2001}. The dashed black line separates Seyferts (left of the line) and LINERs (right of the line) \citep{Kewley+2006}. These lines are traditionally used in AGN-related studies \citep{Polimera+2022, Salehirad+2022, Mezcua+2024a}. A new demarcation line put forth by \citet{Law+2021} is shown by the dotted black line. They used line-of-sight velocity distributions of MaNGA sources to constrain the dynamical properties of the ionized gas and provided new demarcation lines for all three BPT diagrams.

From Figure~\ref{fig:sii-bpt}, we see that the density of sources in the star-forming branch (dense black region) extends beyond the solid black line. In particular, the density contours on the top-left corner that overlap the star-forming branch extend above the demarcation line by \citet{Kewley+2001} and are instead more consistent with the \citet{Law+2021} demarcation line. This finding suggests that the demarcation line needs to be revised for the \sii-BPT diagnostic. 

By considering the \citet{Kewley+2001} line, we find 55,208/449,295 AGN candidates (AGN Fraction $\approx$ 12.3\%). However, if we consider the \citet{Law+2021} line, we identify 27,953/449,295 AGN candidates (AGN Fraction $\approx$ 6.2\%), nearly half of the traditional diagnostic. When we focus on dwarf galaxies, the AGN fraction changes dramatically from 17.4\% with the \citet{Kewley+2001} line to just 4.6\% with the \citet{Law+2021} line. This stark difference highlights that the AGN fraction is sensitive to the chosen diagnostic line. This could explain the unusually high dwarf AGN fraction in dwarf galaxies reported by \citet{Polimera+2022} compared to other single-fiber spectroscopic studies. Some galaxies near these demarcation lines may exhibit both star formation and AGN activity. Therefore, careful analysis is essential for accurately estimating the AGN fraction using these diagnostic methods. 

To identify AGN signatures in galaxies and facilitate comparisons with similar studies, especially in the dwarf galaxy regime, we utilize the \nii-BPT diagram in this paper, while deferring the analysis of the other two diagnostics for future publications. Even with just this single diagnostic, we identify a robust sample of dwarf AGN candidates (Section~\ref{subsec:bpt_diagram}). Incorporating the other two diagnostics will likely reveal an even higher dwarf AGN fraction than what we observe in this study. 

\subsection{Dwarf AGN Fraction: DESI vs. SDSS} \label{subsec:reines13}

Using the \nii-BPT diagnostic, \citet{Reines+2013} reported a dwarf AGN fraction of $\approx$0.5\% based on SDSS DR8 spectroscopic data, which utilized the original SDSS spectrograph. Applying the same diagnostic on the early DESI data, our study reveals a dwarf AGN fraction of $\approx$2.1\% (Section~\ref{subsec:dwarf_agn}), nearly four times higher than this previous estimate. We investigate possible explanations for this increased fraction, such as differences in target selection or instrumental characteristics (e.g., aperture size, spectral resolution). We begin by comparing the targets from both surveys, followed by building a matched sample between SDSS and DESI to isolate the impact of the instrument design. 

\begin{figure}
     \centering
     \includegraphics[width = 0.9\columnwidth]{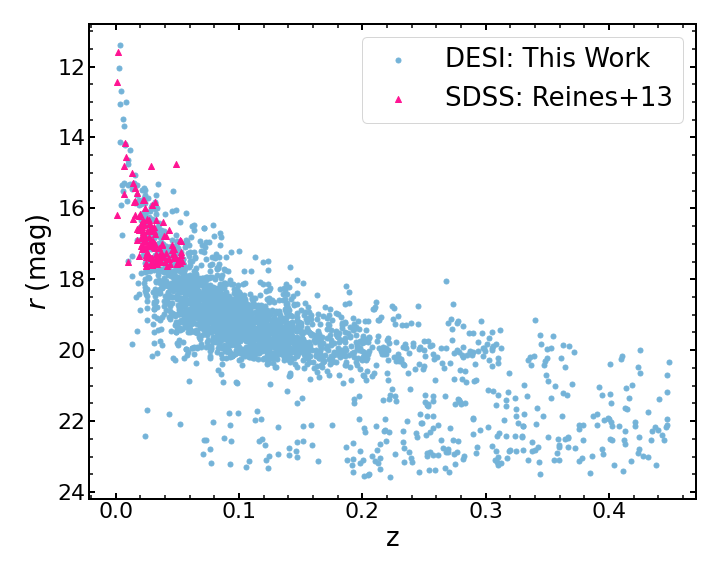}
     \caption{Distribution of dwarf AGN candidates from this work (shown as blue circles) and from \citet{Reines+2013} (shown as pink triangles) in the $r~-$ $\rm z$ space. Our sample of dwarf AGN candidates extend down to fainter magnitudes and to higher redshifts compared to the ones identified by \citet{Reines+2013}.}
     \label{fig:r_z}
\end{figure}

The DESI targets probe fainter sources than the SDSS targets. Figure~\ref{fig:r_z} compares dwarf AGN candidates from our study (shown as blue circles) with those identified by \citet{Reines+2013} (shown as pink triangles) in the $r~-$ $\rm z$ space. Our candidates reach approximately 2.5$-$5 magnitudes fainter and extend about 10 times further in redshift compared to the \citet{Reines+2013} sample. 

To facilitate a direct comparison, we create matched samples of galaxies with similar redshift and $r$ magnitudes to compare DESI vs. SDSS spectra. For this purpose, we first recreate the starting sample of \citet{Reines+2013} using the NASA-Sloan Atlas (NSA). By applying the same selection criteria outlined in their paper, we identify 24,966 dwarf galaxies\footnote{This number differs slightly from the 25,974 sources reported by \citet{Reines+2013}}. We cross-match these galaxies with the LS DR9 catalog to establish a common photometric selection, focusing on sources brighter than the SDSS magnitude limit of $r < 17.7$ mag. We obtain 21,468 dwarf galaxies including 123 candidate AGN, resulting in an AGN fraction of $\approx$0.6\% based on the SDSS spectra, in agreement with \citet{Reines+2013}.

We check the available sample size spanning the same parameter space of $r < 17.7$ mag and $\rm z \leq 0.055$. The number of DESI targets (4,999 sources) within those limits is smaller than the SDSS targets (21,468 sources). Consequently, we identify up to three potential SDSS sources for each DESI target by locating the nearest neighbors in the $r -\rm z$ space. Some DESI sources have only one closest neighbor, resulting in 4,798 unique DESI sources with a total of 10,478 SDSS neighbors. 

To validate our matched sample, we perform the Kolmogorov $-$ Smirnov (K-S test) on $r$ and $\rm z$ distributions of these sub-samples, yielding a $p$-value of 0.99 and 1.0, respectively. These high values indicate that the matched samples have similar distributions in both $r$ and $\rm~z$. We identify 189/4,798 dwarf AGN candidates (AGN Fraction $\approx$3.9\%) from DESI and 64/10,478 dwarf AGN candidates (AGN Fraction $\approx$0.6\%) from SDSS. This difference in AGN fraction is even more pronounced, increasing from a factor of 4 to nearly 8 for the matched DESI sample compared to the SDSS-matched sample. This suggests that the DESI target selection does not account for the observed higher AGN fraction. 

We now focus on the differences in instrument properties. The DESI spectrograph offers higher spectral resolution (60 $-$ 150 $\rm km s^{-1}$) compared to the SDSS spectrograph (100 $-$ 190 $\rm km s^{-1}$). We note that the SDSS resolution is already sufficient to isolate the emission lines used in the BPT diagram and to resolve the profile of \ha\ in case of broad components. Nevertheless, it is plausible that a higher spectral resolution can lead to a more efficient detection of faint emission lines. Additionally, DESI fibers have a smaller diameter (1.5$\arcsec$) compared to SDSS fibers (3$\arcsec$). Since AGNs are typically located at the centers of galaxies\footnote{However, some studies using the MaNGA IFU have reported AGN ionization signatures from wandering and off-nuclear BHs that are often missed by single-fiber spectroscopy \citep[e.g.][]{Wylezalek+2018, Mezcua+2020, Mezcua+2024a}.}, the smaller fiber size of DESI captures light primarily from this central AGN, thereby reducing contamination from the surrounding galaxy and minimizing dilution from star formation \citep{Trump+2015}. We examine these scenarios by comparing the spectra obtained for the same galaxies from the two surveys. 

\begin{figure}
    \centering
    \includegraphics[width=1.0\columnwidth]{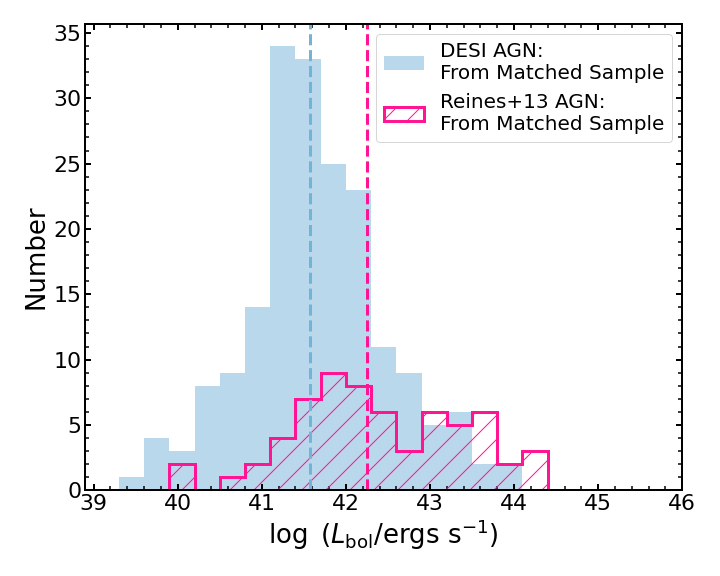}
    \caption{Distribution of AGN bolometric luminosities of the dwarf AGN candidates from the matched starting sample between DESI and SDSS. The filled blue histogram shows the dwarf AGN candidates from DESI, while the hatched pink histogram shows the dwarf AGN candidates from SDSS. The vertical blue and pink lines denote the median $\log~(L_{bol}/\rm ergs~s^{-1})$ values of these distributions. DESI identifies lower luminosity AGN candidates compared to SDSS in the same $r$ magnitude and redshift range.}
    \label{fig:lbol-matched}
\end{figure}

We perform a positional cross-match between the 21,468 dwarf galaxies from SDSS and our starting sample of dwarf galaxies and find 1,725 common candidates. Among these, we find 9 dwarf AGN candidates from \citet{Reines+2013}, while our analysis reveals a total of 24 dwarf AGN candidates. Of the 9 AGN candidates from SDSS, five are also classified as AGN by our selection, two are classified as star-forming, and two do not meet our SNR criteria. In addition to the five common dwarf AGN candidates, we identify 19 additional candidates not selected as AGN by \citet{Reines+2013}. We visually inspect the DESI spectra and fits of these 19 candidates and confirm that they are valid AGN candidates. Given that the SDSS spectra still detect the emission lines (even if the sources lie on the star-forming branch of the BPT), we conclude that spectral resolution or depth is unlikely to be the primary factor. Instead, we suspect that the smaller fiber size of DESI plays a significant role in detecting a higher number of dwarf AGN candidates compared to SDSS, thereby contributing to the observed increase in the dwarf AGN fraction.

In fact, \citet{Moustakas+2010} demonstrated that the fraction of AGN in a galaxy sample is strongly influenced by the integrated light collected within the spectroscopic aperture (see Table 5 and Figure 5). More recently, \citet{Alban+2023} also found that the number of detected AGN candidates decreases as the aperture size increases. If our supposition is correct, we further expect that a smaller aperture would facilitate the detection of less luminous AGNs by effectively isolating their emission from those of the surrounding star-forming regions.

We estimate the bolometric luminosities ($L_{bol}$) of all dwarf AGN candidates from the matched samples using the \oiii~luminosity, following the formula ($L_{bol} = L_{\oiii}\times 1000$) from \citet{Moran+2014}. For the 64 candidates from SDSS, the $\log~(L_{bol}/\rm ergs~s^{-1})$ values range from 40.0 $-$ 44.2, with a median of 42.2. In contrast, the $\log~(L_{bol}/\rm ergs~s^{-1})$ values for the 189 DESI dwarf AGN candidates span from 39.3 $-$ 44.0, with a median of 41.6, $\approx$0.6 dex lower than the SDSS candidates. 

Figure~\ref{fig:lbol-matched} shows the distribution of the $\log~(L_{bol}/\rm ergs~s^{-1})$ values for DESI (in blue) and SDSS (in pink). The lower number of DESI sources at higher luminosities can be attributed to the 1:3 selection ratio of the matched sample. However, as the figure indicates, DESI recovers a greater overall number of AGN candidates compared to SDSS. We perform the K-S test on these luminosities and obtain a $p$-value of $\rm\approx 1.4 \times 10^{-5}$, suggesting that the luminosity distributions for the two samples are significantly different. This indicates that DESI effectively probes lower luminosity AGN candidates compared to SDSS within the same $r~-$ $\rm z$ range, further supporting our hypothesis that the aperture size is indeed the dominant effect driving the difference in the dwarf AGN fraction.

\subsection{BH Seed-Formation Mechanisms}\label{subsec:bh_seeds}

The presence of AGN in dwarf galaxies and the Intermediate-Mass Black Holes (IMBHs) that power them hold clues to the origin of SMBHs in the early Universe. In this subsection, we explore our results of the $\mbh - \mstar$ scaling relation in the context of available literature regarding BH seed formation models.

Early BH seeds can be broadly divided into two types depending on their initial mass \citep{Mezcua2017, Greene+2020}: 1) {\it Light} seeds ($\mbh \approx~10^{2} - 10^{3}~\msun$) that form as end products of Population III stars in the early Universe (``Pop-III scenario'');  2) {\it Heavy} seeds ($\mbh \approx~10^{4} - 10^{6}~\msun$) that form when pristine gas in the early Universe directly collapses into a BH instead of undergoing fragmentation to form stars (``Direct Collapse scenario''). As we cannot directly observe their formation with our current telescopes, we focus on their imprints in the local universe. Most of these early BHs grow via accretion and mergers into the SMBHs we see today. But, a fraction of these might be left over in isolated galaxies and we expect them to be observed as IMBHs in local dwarf galaxies.  

Theoretical studies propose that the low-mass end of the $\mbh - \sigma_{\star}$ relation can help differentiate between these BH seed formation models \citep{Volonteri+2009, Volonteri2010}. While these studies specifically examine the correlation between the BH mass and stellar velocity dispersion, the expected correlations are likely similar to the relationship between BH mass and stellar mass. If the initial BHs are heavy seeds, the scaling relation is predicted to flatten at the low-mass end. Conversely, if the initial seed BHs are light seeds, the scaling relation from higher-mass galaxies is expected to extend down to lower masses. This is based on the assumption that the observed black holes in local dwarf galaxies represent remnant seed black holes from the early universe, which experienced minimal evolution. 

From Figure~\ref{fig:mbh-mstar}, we observe that our scaling relation extends down to $\logmass \approx$ 8.5, with no evident flattening in the relation. The spread among low-mass galaxies appears similar to that of high-mass galaxies. Given that our candidates reach down to down to $\logmbh \approx$ 4.4, with most residing below the established relation, these results may suggest a Population-III seed-formation scenario. Despite their uncertain stellar masses, incorporating the point sources from Figure~\ref{fig:mbh-mstar-morph} could provide additional insights into this relation. If the stellar masses of these sources are accurate, their distribution is consistent with our confident candidates. Also, if their stellar masses are under-estimated, they would shift to the right on the plot, likely aligning with our empirical relation or the one proposed by \citet{Suh+2020}. Conversely, if their stellar masses are over-estimated, they might represent over-massive BHs in dwarf galaxies. In this case, it could lead to a flattening of the $\mbh~-~\mstar$ relation, suggesting a direct collapse seed-formation scenario. 

High-redshift observations using JWST have identified $\gtrsim$20 over-massive BHs pointing to a direct collapse scenario \citep{Harikane+2023, Maiolino+2023, Ubler+2023, Greene+2024}. However, these observations are affected by selection biases favoring the brightest AGN and are likely not detecting the faint sources hosting light seed BHs. Moreover, using a new suite of cosmological simulations, \citet{Bhowmick+2024} suggest that it may be possible to grow these high-redshift over-massive BHs from light seeds with efficient mergers.  

\citet{Ricarte+2018} argued that the low-mass end of the $\mbh - \sigma_{\star}$ scaling relation is driven primarily by the accretion modes of BHs and not the seed-formation models. Furthermore, they show that a hybrid seed-formation model with both light and heavy seeds is almost indistinguishable from a light seed model when studying the scaling relations. \citet{Regan+2024} also emphasized that both seed formation scenarios can occur simultaneously. They suggest that rather than a bimodal distribution of light and heavy seeds, there exists a continuum of BH seeds with the light seeds more abundant and the heavier seeds becoming rarer.

The increased census of dwarf AGN candidates from this study over a spread of BH masses provides an avenue to study the accretion modes in these sources. This number will increase with future DESI releases and will begin the era of the statistical study of dwarf galaxies and their central BHs. Multi-wavelength observations, including future releases of eROSITA X-ray data \citep{eROSITA, Sacchi+2024}, will be useful in constraining the low-mass end of the galaxy-BH scaling relations and their connection to the BH seed formation models. 

\section{Conclusions} \label{sec:conclusions}
Using early DESI spectroscopic data from DESI EDR and 20\% of Year 1 (DA0.2), we identify AGN in a sample of 1,385,177 galaxies split into dwarf ($\logmass \leq$ 9.5) and high-mass ($\logmass >$ 9.5) galaxies. We also constrain the $\mbh - \mstar$ scaling relation to lower galaxy and BH masses compared to previous studies. We use the photometric data from LS DR9 to estimate stellar masses with the SED fitting code CIGALE \citep[Section~\ref{subsec:mstar},][]{Siudek+2024}. We develop a Python-based emission-line fitting code, {\tt EmFit}, to measure the fluxes and widths of narrow and broad components of various emission lines. By selecting line-emitting galaxies at $\rm z \le$ 0.45 across the stellar mass range, we conclude the following:

\begin{itemize}
    \item Using the optical emission-line \nii-BPT diagnostic, we find 75,928/296,261 ($\approx$25.6\%) high-mass AGN candidates and  2,444/114,496 ($\approx$2.1\%) dwarf AGN candidates. With these sources, we have more than tripled the existing census of optical dwarf AGN candidates (Section~\ref{sec:agn_selection}). 

    \item DESI has significantly expanded the discovery space for dwarf AGN candidates. We have extended the search to $\approx$1.5 dex lower galaxy masses and up to 10 times higher redshifts compared to previous spectroscopic surveys (Figure~\ref{fig:mstar_z}). The full DESI survey, set to be completed by 2026, will extend the sample size to $\gtrsim$10,000 dwarf AGN candidates.

    \item Our estimation of the dwarf BPT-AGN fraction ($\approx$2.1\%) is nearly four times higher than that from a comparable systematic search using SDSS \citep{Reines+2013}. This increase can be primarily attributed to the smaller fiber size of DESI compared to SDSS, which aids with the identification of lower luminosity AGN within the same magnitude and redshift range (Section~\ref{subsec:reines13}). 

    \item We find that the BPT-AGN fraction in line emitting galaxies increases with increasing stellar mass, from $\approx$1.2\% at $\logmass \approx$8.0 to $\approx$93.5\% at $\logmass \approx$11.4. 
    On average, the BPT-AGN fraction slightly decreases with increasing redshift.
    However, these trends are affected by selection effects including stellar mass variations, emission line detection limits, and DESI targeting algorithms (Section~\ref{subsec:agn_frac}).

    \item Of the 410,757 line-emitting galaxies in our sample, 6,185 ($\approx$1.5\%) sources show a broad $\ha$ component. Using the \nii-BPT diagnostics, we have a sample of 4,181 BL-AGN candidates that are likely powered by an AGN (Section~\ref{subsec:bpt_diagram}).  

    \item We estimate the BH masses of the BL-AGN candidates using the flux and width measurements of the broad $\ha$ component. The BH masses extend down to  $\logmbh \approx$ 4.4 for confident BL-AGN candidates and to  $\logmbh \approx$ 4.3 for tentative candidates. Among these, we find 151 confident (and 147 tentative) BL-AGNs have $\mbh \le 10^{6}~\msun$, making this the largest sample of IMBH candidates to date (Section~\ref{subsec:bh_masses}).

    \item We extend the $\mbh - \mstar$ scaling relation down to $\logmass \approx$ 8.5 and $\logmbh \approx$ 4.4 (Figure~\ref{fig:mbh-mstar}). The empirical fit from these sources has a similar slope to \citet{Reines&Volonteri2015}, but a slightly higher normalization (Section~\ref{subsec:mbh_mstar}).

    \item The majority of the tentative candidates lie below the empirical fit, suggesting that these sources host under-massive BHs assuming that their broad lines from BLR kinematics. This could be due to the low efficiency of BH growth in galaxies (Section~\ref{subsec:mbh_mstar}).

\end{itemize}

The anticipated increase in the sample of dwarf AGN candidates over the next five years with DESI will accelerate studies of AGN in dwarf galaxies. This expansion will enhance the identification of AGN candidates at the low-mass end of the $\mbh - \mstar$ scaling relation, thereby providing insights into BH seed formation. The statistical sample of dwarf AGN candidates will be invaluable for addressing several key questions related to galaxy evolution on the smallest scales, including accretion modes in low-mass galaxies and the co-evolution of galaxies and their central BHs. 

\section*{Data Availability}
All the data from the figures are available in machine-readable form at \url{https://doi.org/10.5281/zenodo.14009453}.

\section*{Acknowledgements}
The authors thank the anonymous reviewer for their valuable comments and suggestions that helped improve this paper. R.P. thanks Gurtina Besla, Xiaohui Fan, Rob Kennicutt, and Anil Seth for their valuable comments on this project.

R.P. is currently supported by the University of Utah, and was also previously supported by the University of Arizona and in part by NSF NOIRLab. The research of S.J. and A.D. is supported by the U.S. NSF NOIRLab, which is operated by the Association of Universities for Research in Astronomy (AURA) under a cooperative agreement with the National Science Foundation. M.S. acknowledges support by the Polish National Agency for Academic Exchange (Bekker grant BPN/BEK/2021/1/00298/DEC/1), the State Research Agency of the Spanish Ministry of Science and Innovation under the grants `Galaxy Evolution with Artificial Intelligence' (PGC2018-100852-A-I00) and `BASALT' (PID2021-126838NB-I00).  This work was partially supported by the European Union's Horizon 2020 Research and Innovation program under the Maria Sklodowska-Curie grant agreement (No. 754510). M.M. acknowledges support from the Spanish Ministry of Science and Innovation through the project PID2021-124243NBC22. This work was partially supported by the program Unidad de Excelencia Mar\'ia de Maeztu CEX2020-001058-M. 

This material is based upon work supported by the U.S. Department of Energy (DOE), Office of Science, Office of High-Energy Physics, under Contract No. DE–AC02–05CH11231, and by the National Energy Research Scientific Computing Center, a DOE Office of Science User Facility under the same contract. Additional support for DESI was provided by the U.S. National Science Foundation (NSF), Division of Astronomical Sciences under Contract No. AST-0950945 to the NSF’s National Optical-Infrared Astronomy Research Laboratory; the Science and Technology Facilities Council of the United Kingdom; the Gordon and Betty Moore Foundation; the Heising-Simons Foundation; the French Alternative Energies and Atomic Energy Commission (CEA); the National Council of Humanities, Science, and Technology of Mexico (CONAHCYT); the Ministry of Science, Innovation, and Universities of Spain (MICIU/AEI/10.13039/501100011033), and by the DESI Member Institutions: \url{https://www.desi.lbl.gov/collaborating-institutions}. Any opinions, findings, and conclusions or recommendations expressed in this material are those of the author(s) and do not necessarily reflect the views of the U. S. National Science Foundation, the U. S. Department of Energy, or any of the listed funding agencies.

The authors are honored to be permitted to conduct scientific research on Iolkam Du’ag (Kitt Peak), a mountain with particular significance to the Tohono O’odham Nation.

The DESI Legacy Imaging Surveys consist of three individual and complementary projects: the Dark Energy Camera Legacy Survey (DECaLS), the Beijing-Arizona Sky Survey (BASS), and the Mayall z-band Legacy Survey (MzLS). DECaLS, BASS and MzLS together include data obtained, respectively, at the Blanco telescope, Cerro Tololo Inter-American Observatory, NSF’s NOIRLab; the Bok telescope, Steward Observatory, University of Arizona; and the Mayall telescope, Kitt Peak National Observatory, NOIRLab. NOIRLab is operated by the Association of Universities for Research in Astronomy (AURA) under a cooperative agreement with the National Science Foundation. Pipeline processing and analyses of the data were supported by NOIRLab and the Lawrence Berkeley National Laboratory (LBNL). Legacy Surveys also uses data products from the Near-Earth Object Wide-field Infrared Survey Explorer (NEOWISE), a project of the Jet Propulsion Laboratory/California Institute of Technology, funded by the National Aeronautics and Space Administration. Legacy Surveys was supported by: the Director, Office of Science, Office of High Energy Physics of the U.S. Department of Energy; the National Energy Research Scientific Computing Center, a DOE Office of Science User Facility; the U.S. National Science Foundation, Division of Astronomical Sciences; the National Astronomical Observatories of China, the Chinese Academy of Sciences and the Chinese National Natural Science Foundation. LBNL is managed by the Regents of the University of California under contract to the U.S. Department of Energy. The complete acknowledgments can be found at \url{https://www.legacysurvey.org/acknowledgment/}.

This research uses services and data provided by the Astro Data Lab, which is part of the Community Science and Data Center (CSDC) program at NSF NOIRLab.

\facilities{DESI, Mayall}

\software{Astropy \citep{astropy+2013, astropy+2018, astropy+2022}, CIGALE \citep{cigale}, FastSpecFit \citep{fastspecfit}, Matplotlib \citep{matplotlib}, NumPy \citep{numpy}, QuasarNet \citep{quasarnet}, Redrock \citep{redrock_qso, Anand+2024}}


\bibliography{references}{}
\bibliographystyle{aasjournal}



\appendix
\section{Double-Peaked Emission Line Sources}\label{app:double-peak}

\begin{figure*}[ht!]
    \centering
    \includegraphics[width=1.0\textwidth]{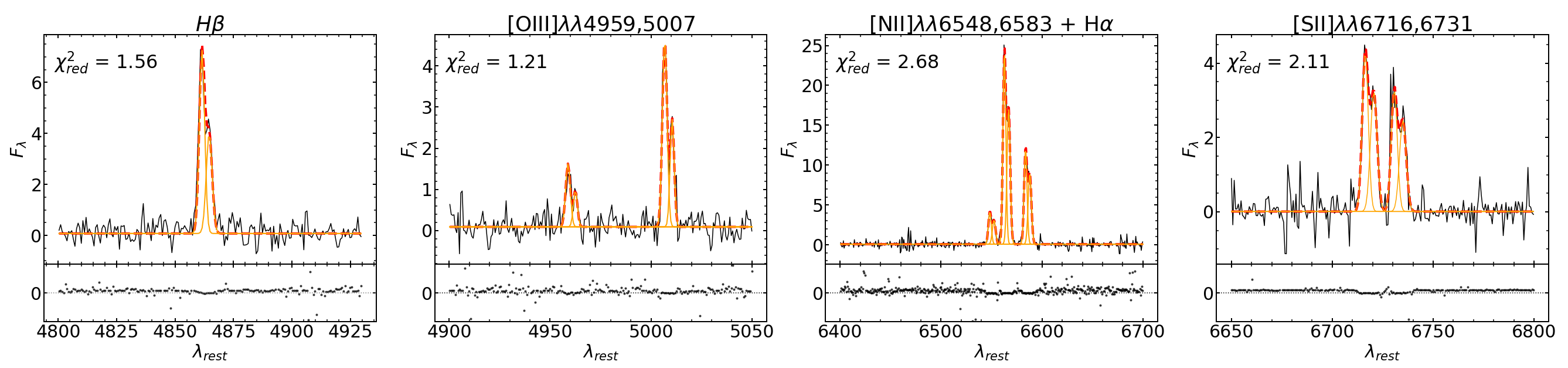}
    \caption{Best-fit model example of a double-peaked emission line candidate (DESI J86.0033-22.4275): The best-fit models to the continuum-subtracted emission-line spectrum in the regions - $\hb$, \oiii, \nii~+ $\ha$, and \sii~- from left to right. The spectrum is shown in black, while the best-fit models are shown in dashed red. The individual narrow and extra components are plotted in orange. The reduced $\chi^{2}$ values for each fit are given in the upper left corner of the individual panels. The fractional residuals are plotted as gray points in the bottom panels for the fits. This galaxy requires two components to fit all the narrow emission lines; however, most galaxies classified as double-peaked have more ambiguous line profiles and/or a second peak apparent for only a subset of emission lines.}
    \label{fig:double-peak}
\end{figure*}

When fitting narrow emission lines (\sii~and \oiii), we evaluate the presence of additional components as described in Section~\ref{subsec:emfit}. These extra components may correspond to an outflow component (see Figure~\ref{fig:default-mode}) or an additional narrow component. Whenever an extra component is detected in \sii, we apply the same profile to \nii, $\ha$, and $\hb$ fits. 

Figure~\ref{fig:double-peak} shows an example of a double-peaked emission line candidate that requires two components to fit all the narrow lines: $\hb$, \oiii, \nii, $\ha$, and \sii. Such dual-peaks in emission lines can be a result of complex gas kinematics (e.g., double nuclei, accretion disk kinematics, ejecta, bar) in the galaxy \citep[e.g.,][]{Maschmann+2023}.  However, not all galaxies exhibit a resolved second component across all narrow lines simultaneously. In fact, of the 26,819 galaxies in our sample that show a detectable second component in at least one of the four emission lines used in the \nii-BPT diagram, only 2,803 galaxies have a non-zero flux in the second component across all four emission lines. To ensure accurate AGN classification, it is therefore important to identify these sources and carefully account for their emission line fluxes.

\begin{figure*}[ht!]
    \centering
    \includegraphics[width=1.0\textwidth]{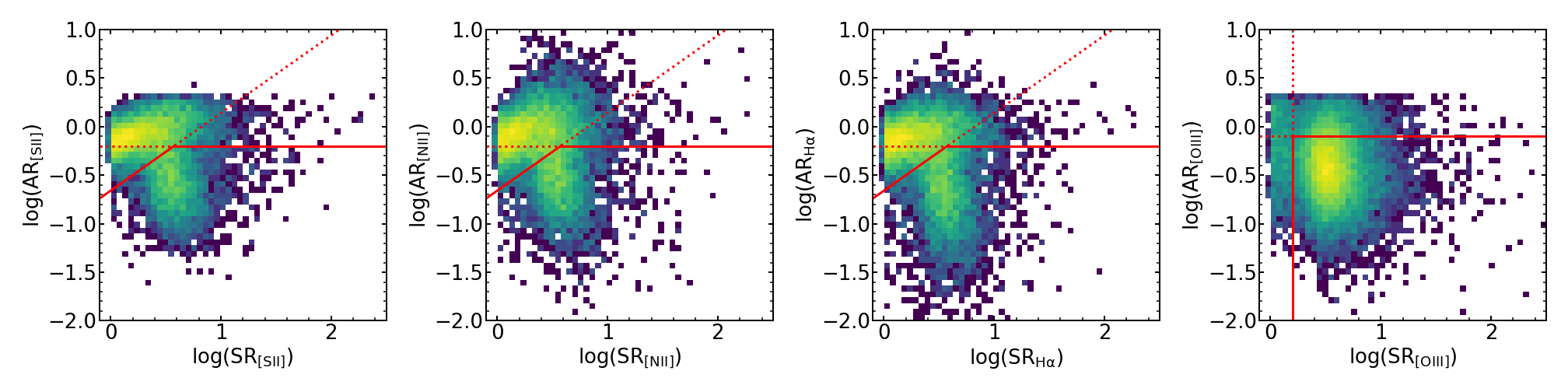}
    \caption{Bivariate distribution of sources in the Amplitude ratio - $\sigma$ ratio space for \sii, \nii, $\ha$, and \oiii~- from left to right. Only the sources with two components from the fitting method are plotted. The solid red lines denote the selection criteria for separating the fits with outflow components from the fits with extra narrow components.}
    \label{fig:doublepeak-separation}
\end{figure*}

\begin{figure*}
    \centering
    \includegraphics[width=1.0\textwidth]{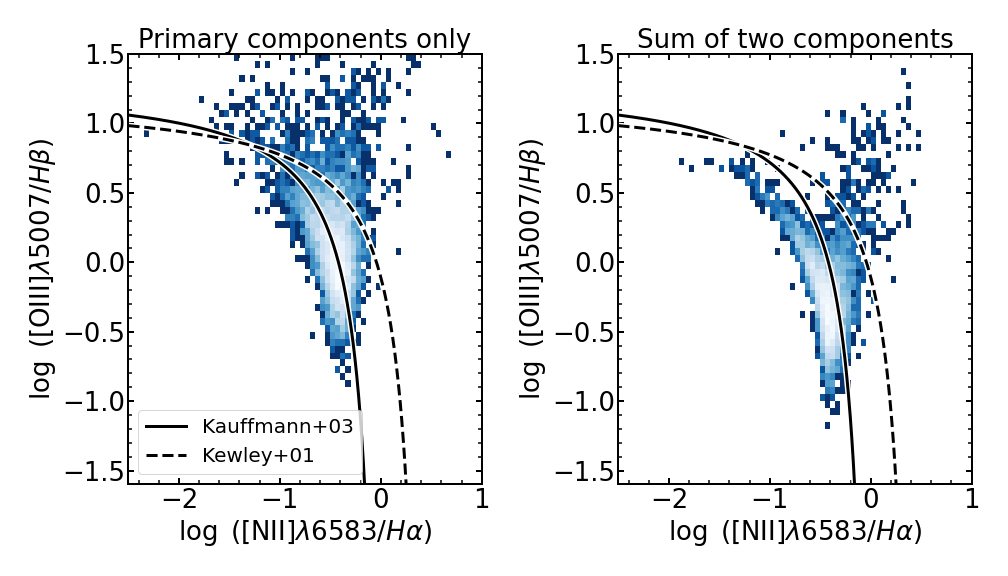}
    \caption{\nii-BPT diagram of the 6,230 sources that exhibit an extra component (which is not an outflow component) in \sii, \nii, $\ha$, and $\hb$, but show no extra component in \oiii. {\it Left: } Emission-line ratios are estimated using only the primary components. {\it Right:} Emission-line ratios are estimated by summing the fluxes of the two components in each emission line.}
    \label{fig:bpt-comparison}
\end{figure*}

For the double-peaked emission lines, we find that either the amplitudes or the widths of the two components are similar. This is in contrast with the outflows for which the width and amplitude of the two components are distinct. We define the amplitude ratios and $\sigma$ ratios of the two components for each of the emission lines as follows:

\begin{equation*}
    \rm AR_{em} = \frac{Amplitude_{em;out}}{Amplitude_{em}}
\end{equation*}
\begin{equation*}
    \rm SR_{em} = \frac{\sigma_{em;out}}{\sigma_{em}}
\end{equation*}

\noindent where `em' denotes the different emission lines: \sii, \nii, $\ha$, and \oiii. The second component ($\rm em;out$) is defined such that its width is always greater than the width of the primary component, leading to $\rm SR_{em}$ being always greater than 1.

We identify 10,934 sources that show an extra component in \sii~emission line fit, where the same profile is then used to fit \nii, $\ha$, and $\hb$ emission lines. In Figure~\ref{fig:doublepeak-separation}, we show the distribution of these sources in the $\rm \log(AR)~-~\log (SR)$ space for \sii, \nii, and $\ha$ lines. We see that there are two different distributions of sources in these three panels. The cases where the amplitude of the extra component is greater than that of the primary component ($\rm\log(AR) > 0$) are sources without an outflow component. We visually inspect the sources where either the amplitude or $\sigma$ values of the two components are nearly equal. We focus on \nii~and $\ha$ lines as they are critical for the \nii-BPT diagnostic. Based on the visual inspection, we find the following optimum cut for separating the double-peaked emission line sources from the outflow candidates:

\begin{equation}
    \rm \log (AR_{[NII]}) > ((0.8\log (SR_{[NII]})) - 0.66)~OR~\log (AR_{[NII]}) > -0.2
    \label{eq:a1}
\end{equation}
\begin{center}
    OR
\end{center}
\begin{equation}
    \rm \log (AR_{\ha}) > ((0.8\log (SR_{\ha})) - 0.66)~OR~\log (AR_{\ha}) > -0.2
    \label{eq:a2}
\end{equation}

This criterion is shown as solid red lines in the left three panels of Figure~\ref{fig:doublepeak-separation}. All the sources above these solid red lines (7,789/10,934) are the cases where the extra component is not considered as an outflow component. 

We repeat the visual inspection for the \oiii~emission line. The right-most panel of Figure~\ref{fig:doublepeak-separation} shows the distribution of 18,725 sources that have an extra \oiii~component, in the $\rm \log(AR)~-~\log (SR)$ space. From visual inspection, we find the following optimum cut for the \oiii~components:

\begin{equation}
    \rm \log (AR_{[OIII]}) > -0.09~OR~\log (SR_{[OIII]}) < 0.2
    \label{eq:a3}
\end{equation}

The solid red lines in the \oiii~panel of Figure~\ref{fig:doublepeak-separation} mark this criterion and the sources that are above or left of these lines (4,247/18,725) are the cases where the extra component is not considered as an outflow component. 

We classify these sources selected via the above criteria as double-peaked emission-line candidates, and they have several complexities associated with their emission-line flux measurements. The dual peaks are only marginally resolved and are not consistently present across all emission lines. In several sources, there is also some ambiguity in identifying the primary component of the emission line. For a robust analysis of their ionization signatures, we sum the fluxes of the individual components to derive the total emission-line flux. The uncertainties in these fluxes are estimated by adding the uncertainties of the individual components in quadrature.

To illustrate the importance of this approach, we analyze the \nii-BPT diagram of the 6,230 galaxies with double peaks in \nii, $\ha$, and $\hb$ emission lines, but no extra component in \oiii. We compare the results when using only the primary component (left panel of Figure~\ref{fig:bpt-comparison}) versus when summing both the components (right panel of Figure~\ref{fig:bpt-comparison}). We find that the star-forming and AGN branches are unclear when using the primary components alone, but are visible when using the sum of the two components. The left panel of Figure~\ref{fig:bpt-comparison} results in 3,255 sources ($\approx$52\%) that lie on the AGN-dominated or composite region of the BPT diagram. On the other hand, the right panel results in only 1,288 sources ($\approx$21\%) lying on these regions of the BPT diagram. This suggests that relying solely on the primary components leads to an over-detection of AGN candidates. We, therefore, adopt this procedure for all the emission line-related studies mentioned in this paper (Sections~\ref{sec:methods} and \ref{sec:agn_selection}).

\section{Visual Inspection of Broad-Line Candidates}\label{app:vi}  
Given that we are extending the search for broad components of $\ha$ to narrower widths (FWHM ($\ha$;b) $\rm > 300~km~s^{-1}$), we want to ensure that we are including robust BL-AGN candidates in our analysis. We therefore visually inspect the spectra and fits of 556 BL sources with (FWHM ($\ha$;b) $\rm < 1000~km~s^{-1}$) and which are BPT-AGN or BPT-composite candidates, and assign them one of the four VI Flags. These are not to be confused with the standard VI Flags used for DESI Survey Validation \citep{Alexander+2023}. 

\begin{figure*}[ht!]
    \centering
    \includegraphics[width = 1.0\textwidth]{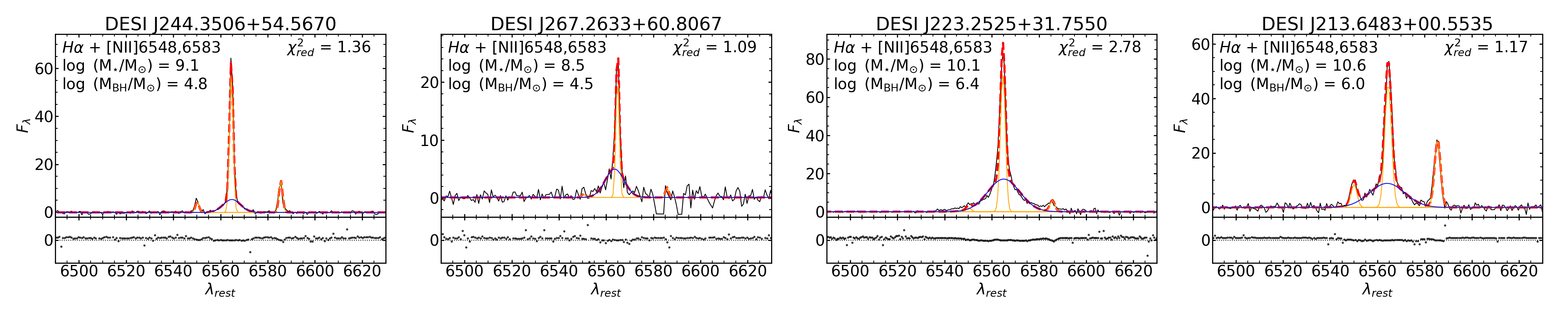}
    \caption{Example emission-line spectra and fits of the \nii~+~$\ha$ region of four candidates with VI Flag = 0 (confident BL-AGN candidates). The spectrum is shown in black, while the best-fit model is shown in dashed red. Individual narrow and outflow components are plotted in orange, and the broad component is shown in blue. The reduced $\chi^{2}$ values for these fits are shown in the upper right corner of the panels. Additionally, the stellar mass of the galaxy and the estimated BH mass derived from the broad component are noted in the top left corner of the panels. These are sources with visually clear broad $\ha$ components.}
    \label{fig:vi0}
\end{figure*}

\begin{itemize}
    \item {\bf VI Flag = 0:} This flag denotes good fits with clear broad components. We consider these as {\it confident} BL-AGN candidates. Of the 556 sources, we have 146 sources that satisfy this visual confidence. Figure~\ref{fig:vi0} shows example cases of four such candidates.

    \begin{figure*}[ht!]
        \centering
        \includegraphics[width = 1.0\textwidth]{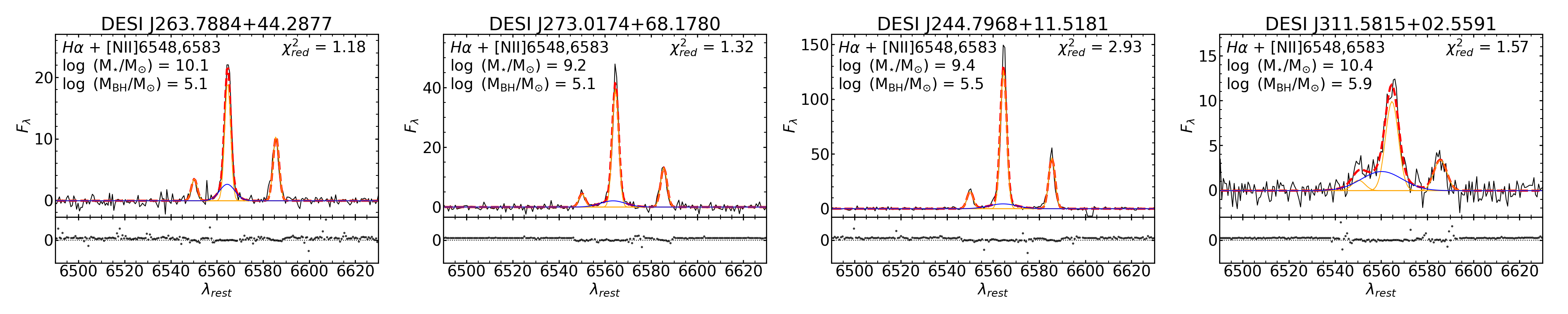}
        \caption{Example emission-line spectra and fits of the \nii~+~$\ha$ region of four candidates with VI Flag = 1 ({\it tentative} BL-AGN candidates). The color coding of the curves and the text in the panels matches are the same as Figure~\ref{fig:vi0}. These are sources with visually unclear, yet statistically detected broad $\ha$ component.}
        \label{fig:vi1}
    \end{figure*}

    \item {\bf VI Flag = 1:} This flag denotes cases where a broad $\ha$ component is detected statistically but is not obvious in visual inspection. Of the 556 sources, we have 154 such cases and we call them {\it tentative} BL-AGN candidates and include them in our analysis (Section~\ref{sec:scaling_relation}). Figure~\ref{fig:vi1} shows four example candidates with this flag.

    \item {\bf VI Flag = 2:} This flag is for candidates where the fitting code fails to identify a second or outflow component. This issue arises when the \sii~region is noisy and the fitting does not detect the second or outflow component, but is visible in the \nii$\lambda$6583 line. Consequently, the broad $\ha$ component in these cases may correspond to the missed second or outflow component, rather than being attributed to the kinematics of the BLR. As a result, the estimated BH masses for these candidates are not accurate, and we exclude 252/556 sources with this VI flag. Four examples of such cases are illustrated in Figure~\ref{fig:vi2}.

    \begin{figure*}[ht!]
        \centering
        \includegraphics[width = 1.0\textwidth]{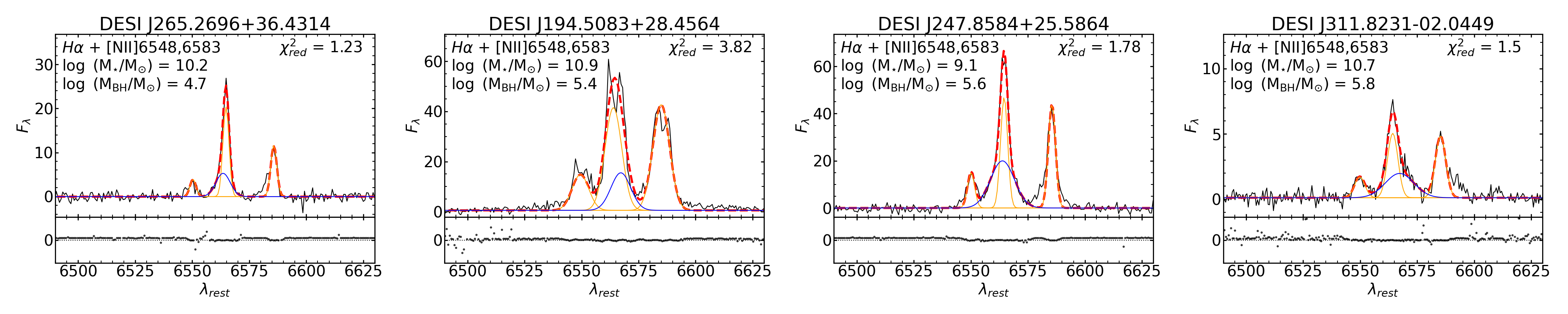}
        \caption{Example emission-line spectra and fits of the \nii~+~$\ha$ region of four candidates with VI Flag = 2.  The color coding of the curves and the text in the panels matches are the same as Figure~\ref{fig:vi0}. These are sources where the broad $\ha$ component is possibly a missed second or outflow component that is visually clear from the \niilam~emission line. The indicated BH mass is estimated under the assumption that the broad component originates from the virialized gas near the BH; however, this assumption may not be valid in this case.}
        \label{fig:vi2}
    \end{figure*}

    \begin{figure*}[ht!]
        \centering
        \includegraphics[width = 1.0\textwidth]{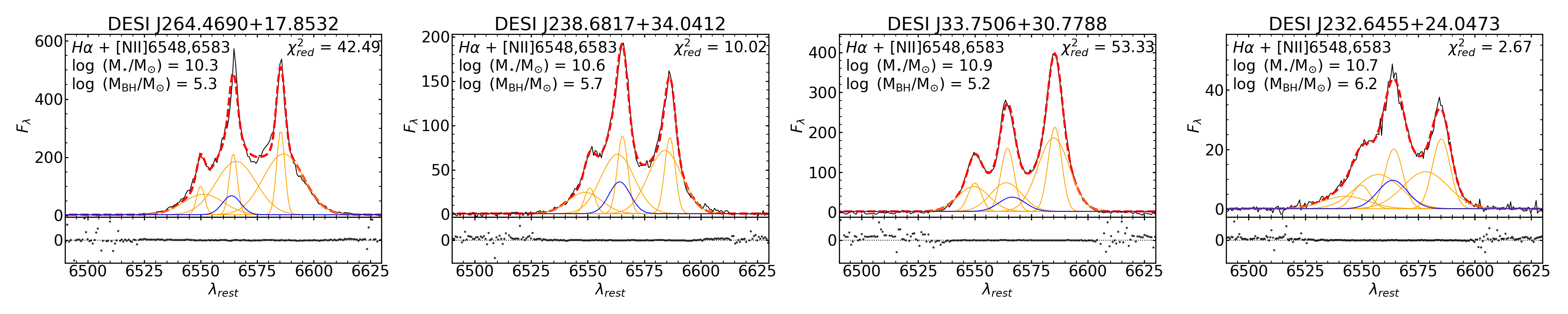}
        \caption{Example emission-line spectra and fits of the \nii~+~$\ha$ region of four candidates with VI Flag = 3. The color coding of the curves and the text in the panels matches are the same as Figure~\ref{fig:vi0}. These are sources where it is challenging to accurately separate the broad $\ha$ component due to complex kinematics. The indicated BH mass is estimated under the assumption that the broad component originates from the virialized gas near the BH; however, this assumption may not be valid in this case.}
        \label{fig:vi3}
    \end{figure*}
    
     \item {\bf VI Flag = 3: } This flag denotes cases with uncertain fits due to complex kinematics in the \ha\ and \nii\ region, with only four candidates falling into this category (see Figure~\ref{fig:vi3}). All of these sources exhibit strong radio emission, as observed in VLA Sky Survey images available through the LS Sky Viewer\footnote{\url{https://www.legacysurvey.org/viewer}}. While this suggests the presence of an AGN, the broad $\ha$ components lead to estimates of under-massive black holes. The kinematics in these cases may be complex, making it challenging for {\tt EmFit} to accurately separate the emission components. We therefore exclude these candidates from our analysis.
\end{itemize}

\begin{figure*}
        \centering
        \includegraphics[width = 0.8\textwidth]{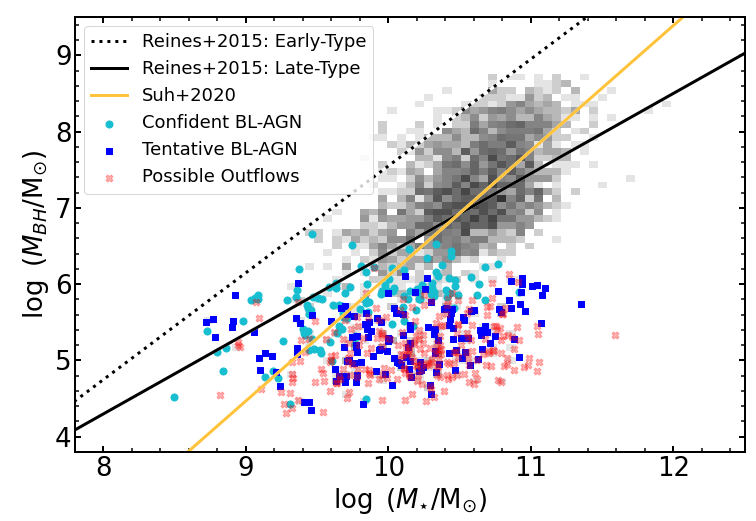}
        \caption{Position of visually inspected BL-AGN candidates on the $\mbh - \mstar$ space. The confident (VI Flag = 0), tentative (VI Flag = 1), and outflow (VI Flag = 2) candidates are shown as cyan circles, blue squares, and red crosses, respectively. }
        \label{fig:mbh_mstar_vi}
    \end{figure*}

In Figure~\ref{fig:mbh_mstar_vi}, we show the location of these visually inspected BL-AGN candidates on the $\mbh - \mstar$ space based on their VI flag. The confident (VI Flag = 0) and tentative candidates (VI Flag = 1) are shown as cyan circles and blue squares, respectively. The sources where the outflow components are possibly selected as broad components (VI Flag = 2) are shown as red crosses. Even though the assumption of BLR kinematics is not valid in such a case, we estimate their BH masses using this extra component. The sources with possible outflows are located in the same region as tentative candidates and are removed from our bestfit estimates of the empirical relation in Section~\ref{subsec:mbh_mstar}. 

\suppressAffiliationsfalse
\allauthors
\end{document}